\documentclass{aastex631}
\usepackage{hhline,colortbl}
\usepackage{amsmath}
\usepackage{placeins}
\usepackage{rotating}

\graphicspath{{./}{figures/}}
\usepackage{svg}
\usepackage{amssymb}
\shorttitle{Nova V1716 Scorpii}
\shortauthors{Worley et al.}
\begin{document}
\title{X-ray observations of Nova Scorpii 2023 (V1716 Sco) in outburst}
\correspondingauthor{John Worley}
\email{jackworley10@gmail.com}
\author[0009-0005-1997-6980]{John T. Worley}
\affil{University of Wisconsin-Madison Astronomy Department \\
475 N Charter St. Madison, Wisconsin, 53706 USA}
\author[0000-0003-1563-9803]{Marina Orio}
\affiliation{University of Wisconsin-Madison Astronomy Department \\
475 N Charter St. Madison, Wisconsin, 53706 USA} 
\affiliation{INAF-Osservatorio di Padova, vicolo Osservatorio 5, I-35122 Padova, Italy}

\author[0009-0004-3869-2425]{Andrej Dobrotka}
\affiliation{Advanced Technologies Research Institute, Faculty of Materials Science and Technology in Trnava, \\
Slovak University of Technology in Bratislava, Bottova 25,
917 24 Trnava, Slovakia}
\author[0000-0001-6833-1875]{Jozef Magdolen}
\affiliation{Advanced Technologies Research Institute, Faculty of Materials Science and Technology in Trnava, \\
Slovak University of Technology in Bratislava, Bottova 25,
917 24 Trnava, Slovakia}

\author[0000-0001-5624-2613]{Kim Page}
\affiliation{School of Physics and Astronomy, University of Leicester, University Road, Leicester, LE1 7RH, UK}

\author[0000-0001-9735-4873]{Ehud Behar}
\affiliation{Department of Physics, Technion, Haifa 32000, Israel}

\author[0000-0002-0210-2276]{Jeremy J. Drake}
\affiliation{Lockheed Martin Solar and Astrophysics Laboratory, 3251 Hanover Street, Palo Alto, CA 94304, USA}
\author[0000-0001-6798-5447]{Sharon Mitrani}
\affiliation{Department of Physics, Technion, Haifa 32000, Israel}

%
\begin{abstract}
Nova Scorpii 2023 was first detected as a luminous supersoft X-ray source (SSS) 93 days after outburst and continued emitting soft X-rays for over two months, until it was too close to the Sun to observe. The nova was monitored with the {\sl Swift} X-ray Telescope (XRT) and the Neutron Star Interior Composition Explorer ({\sl NICER}) on the International Space Station, and in long exposures with the {\sl Chandra} High Resolution Camera (HRC) and Low Energy Transmission Grating (LETG) on days 128, 129, and 183-185 after optical maximum. {\sl Swift} detected a rapidly decaying SSS when observations resumed, constraining the constant bolometric luminosity phase to $\sim$ 9 months. The SSS flux was irregularly variable. A nearly three-fold increase in flux was observed between 2023 August and October in the 15--35 \AA \ range, from 3.5 $\times 10^{-11}$ to 9.4 $\times 10^{-11}$ erg cm$^{-2}$ s$^{-1}$. The SSS duration and effective temperature derived from the October LETG spectra indicate a massive white dwarf with temperature fitting nova evolutionary tracks for a $\approx$ 1.2 M$_\odot$WD; emission lines superimposed on the WD continuum are attributed to surrounding shocked ejecta. We present a timing study based on {\sl Chandra} and archival {\sl NICER} data. The irregular variability timescale was days, but a $\sim$ 77.9 s periodic modulation in the SSS flux with varying amplitude was measured in many observations. Our analysis shows that this period was stable; short drifts derived with {\sl NICER}, but not in long, uninterrupted {\sl Chandra} exposures, are artifacts of measuring variable amplitude modulation. We suggest the modulations are associated with the WD rotation.
\end{abstract}
\section{Introduction} \label{sec:intro}
Novae are highly dynamic outbursts on white dwarfs (WDs) that have been accreting material from a non-degenerate stellar companion; they emit radiation across all wavelengths. In short period systems, the binary usually fills its Roche lobe, causing matter to be accreted by the WD. However, wind driven accretion may be another viable method of mass transfer in long period symbiotic systems \citep[e.g.,][]{Chomiuk21}. The accreted matter builds up on the surface of the WD, becoming electron degenerate and mixing with WD matter - estimates range from 25\% to 50\% of mixed material in the burning layer \citep{Woodward}. Nuclear burning occurs close to the WD surface, until a critical pressure is reached and a thermonuclear runaway (TNR) occurs in the accreted envelope around the star. Timescales for TNR ignition vary, and are mostly dependent on the WD mass and the accretion rate, $\dot{M}_{\text{accr}}$. The mass accretion rate in most novae is estimated to be in the range $\dot{M}_{\text{accr}}$= $10^{-10}$ to $10^{-8} M_{\odot} \text{yr}^{-1}$ \citep{Selvelli,Shara}.
The more massive the WD, the smaller its radius, so the TNR occurs with a lower accreted envelope mass ($\Delta M_{\text{accr}}$) because degeneracy is reached sooner \citep{Yaron}. Novae typically reach a bolometric luminosity of $10^{38}$ $\text{erg } \text{s}^{-1}$ during outburst. Due to the TNR, the envelope expands, and a very fast stellar wind follows, either because of the radiation pressure or following double Roche-lobe filling \citep{ShenQua}, with velocities up to a few thousand km s$^{-1}$\citep{Starrfield2016}. 

As the optical flux declines, the radius of the WD shrinks back to almost pre-outburst dimensions at constant bolometric luminosity. The peak emission shifts towards the UV and eventually to the soft X-ray range (0.1-2.0 keV), beginning the super soft X-ray source (SSS) phase - for the theory see \citet{Yaron2005}, and, for the observations, among other papers, pioneering work done by \citet{Ogelman1984, 2010ApJ...717..363R}. In this phase, the WD atmosphere is observable with an effective temperature in the range of few hundred thousand to a million K. The duration of the SSS phase ranges from a few days to $\simeq$10 years, \citep{2013ApJ...777..136W,2015MNRAS.454.3108P}, depending on the mass accretion rate and WD mass. The SSS temperature depends critically on the WD mass. 

Shortly before the outburst and before the SSS phase, X-ray spectra of novae are characterized by thermal emission of shocked ejecta. Shocks in the outflowing material of novae have been recognized to play an important role in ejecta dynamics. Because of the velocity of the outflow of a few thousand km s$^{-1}$, these shocks cause X-ray emission. The emission from the shocks may continue during the SSS phase, producing an overlapping emission line spectrum in X-rays.

Nova V1716 Sco (PNV J17224490-4137160) was first observed on 2023 April 20.68 by Andrew Pearce at optical magnitude $m_{\text{op}}$ = 8.0\footnote{\url{http://www.cbat.eps.harvard.edu/unconf/followups/J17224490-4137160.html}}. In this work, we adopt this date as T$_{0}$. {\sl Fermi} LAT data from 2023 April 21 show $\gamma$-ray emission from a source coinciding in space and time with V1716 Sco, with a peak energy of 23.5 GeV \citep{2023ATel16002....1C, wang2024}. On 2023 April 22.29, V1716 Sco was confirmed as a classical nova by \citet{2023ATel16003....1W,Chiron_Echelle}, who measured the optical spectrum of a typical, lightly reddened Fe II nova near maximum. The AAVSO optical light curve reached peak magnitude on April 22.6. Optical spectroscopy in the early phases was obtained by \citet{2023ATel16004....1S, 2023ATel16006....1S}, showing high ejection velocities, in the range of 1800-3000 km s$^{-1}$ for the Balmer lines, indicating complex ejecta dynamics. The flux measured with {\sl Fermi} was about two orders of magnitude larger than the flux measured in
 hard X-rays with {\sl NuSTAR} in the 3-50 keV range from April 21.89 to April 23.42 \citep{wang2024}. \cite{wang2024} were not able to discriminate between
  a thermal or non-thermal origin of the flux measured in X-rays; however given the discrepancy in flux, if the X-rays originated in shocked plasma, this shock was not the same that accelerated the particles giving rise to flux in the GeV range, whose X-ray flux was unobservable, presumably because it occurred in a region with extremely high column density along the line of sight.

V1716 Sco was monitored regularly by the {\sl Neil Gehrels Swift Observatory} following the outburst, first detecting X-ray emission on 2023 May 1st, with a very soft X-ray component emerging on May 31st \citep{2023ATel16069....1P}. The evolution of the SSS was followed for 37 days with the Neutron Star Interior Composition Explorer ({\sl NICER}), and \citet{Dethero2023} found
  a quasi-periodic modulation of the SSS with a period close to $\simeq$79 s, later confirmed  by \citet{wang2024}.
\section{X-ray observations of the nova in outburst} \label{sec:Observations}
\begin{figure*}
\centering
\includegraphics[width=0.8\textwidth]{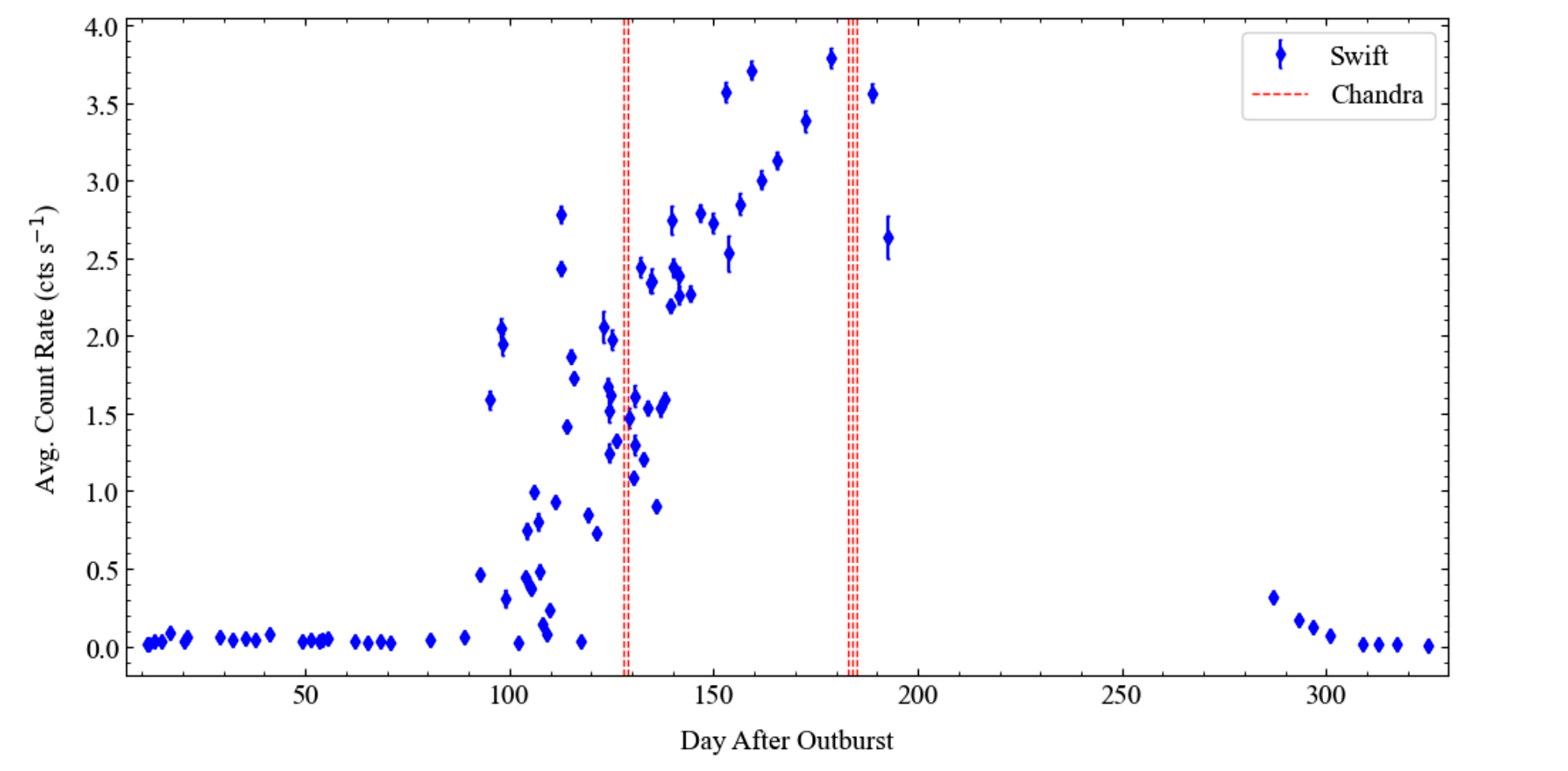}
\includegraphics[width=0.8\textwidth]{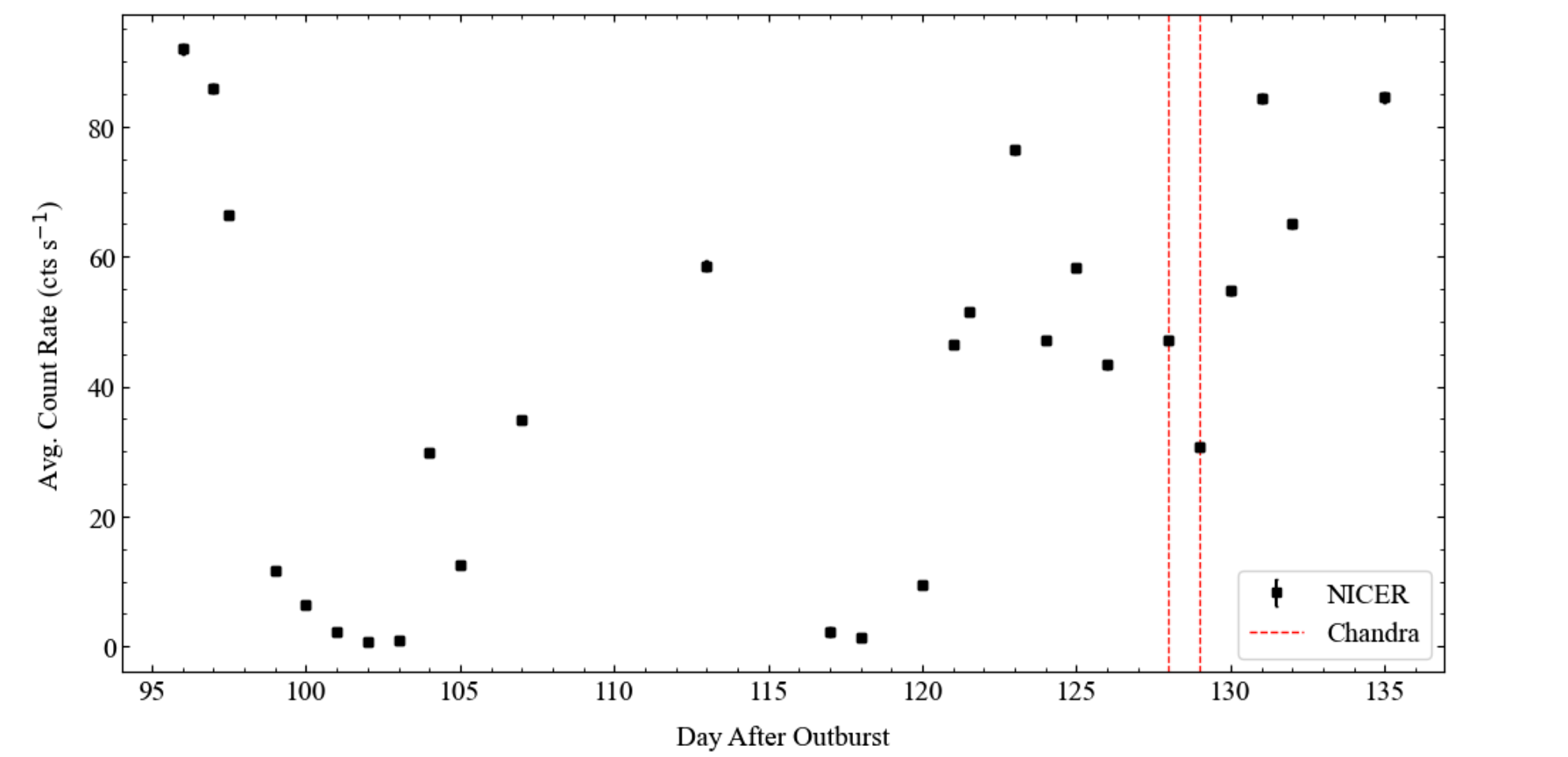}
\caption{Upper panel: light curve of Nova V1716 Sco over 300 days of {\sl Swift} observations in the 0.3-10 keV range. The epochs of the {\sl Chandra} exposures are denoted with dashed red lines.
Lower panel: light curve obtained with {\sl NICER}, with count rate is in the range 0.2-0.8 keV, the energy band in which the flux is significant.}
\label{fig:light_curves}
\end{figure*}

Table~\ref{tab:Swiftdata} shows the dates of the {\sl Swift} observations and the XRT count rates, plotted in Fig.~\ref{fig:light_curves}. 
The {\sl Swift} data were processed using the standard \texttt{xrtpipeline} tool, together with the most recent calibration files available at the time; \texttt{xselect} was then used to extract light curve bins over 0.3-10.0 keV. The last {\sl Swift} observation before the nova was too close to the Sun was performed on 2023 October 30 and the observations resumed only on 2024 February 1st.

From 2023 April 21st until 2023 July 18 the source was faint. We examined the spectra and found that the hardness ratio, defined as the ratio of the count rate between 2 and 10 keV and the count rate between 0.3 and 2 keV, progressively decreased from 10 at the beginning to less than 1 within 90 days. The spectra of the single exposures in 2023 May can be fitted with a model of thermal plasma in collisional ionization equilibrium \citep[APEC in XSPEC, see ][]{Arnaud} with column density above the interstellar value N(H)$\simeq$4.4 $\times$ 10$^{21}$ cm$^{-2}$ \citep{HI4PI2016}, 
 and progressively decreasing temperature. However, for at least the first month, we find that the 3 $\sigma$ confidence
 contours of the plasma temperature versus column density are ``open'' in the portion with high temperature and low column density, so that
 the peak temperature is not constrained by the data. Around optical maximum, on 2023 April 21, the thermal model of \citet{wang2024} used to fit a {\sl NuSTAR} observation
    only constrained the temperature to 20.15$\pm$16.13 keV.

After the detection and subsequent brightening of a luminous supersoft X-ray component, first observed with
 {\sl Swift} in 2023 July 21 three months after optical maximum \citep{2023ATel16150....1P}, 
 28 exposures were taken with {\sl NICER} between 2023 July 25 and 2023 September 2. They are  listed in Table~\ref{tab:NICEROBS} in the Appendix.
 The {\sl NICER} data products, initially presented and examined by \citet{Dethero2023, wang2024},  were retrieved by us from NASA's High Energy Astrophysics Science Archive Research Center (HEASARC).  The {\sl NICER} energy range is 0.2-12.0 keV.   The \texttt{nicerl2} data calibration pipeline was applied to each observation, using the CALDB to provide cleaned event files. After reprocessing, \texttt{nicerl3-lc} and \texttt{nicerl3-spect} with the SCORPEON method to estimate the background were used to create light curve, spectra, background, and response matrix files from the calibrated data. 

 In 2023 August 
 we triggered an approved Target of Opportunity (TOO) proposal to obtain high resolution {\sl Chandra} Low Energy Transmission Grating (LETG) and High Resolution Camera (HRC) spectra. Two exposures were started on 2023 August 26.067 and
  August 27.78, respectively. Three additional exposures started on 2023 October 20.48, 21.32, and 22.63.
 We extracted the five {\sl Chandra} HRC+LETG spectra using the CIAO software package \citep{Fruscione} version
4.15.2 and the CALDB calibration package \citep{Graessle} version 4.10.4. First, the \texttt{chandrarepro} command
was applied to the event file, and subsequently, with the command \texttt{combinegratingspectra}, we averaged the $\pm{1}$
diffraction orders. We extracted the zero
order lightcurves measured with the HRC using CIAO.

The list of observations, with the exposure start time, duration, measured count rate and flux, is in Table~\ref{tab:observations} (Appendix). High spectral resolution grating spectra allow flux measurement, without having to resort to an estimate from fitting a model as when using CCD-type detectors. The flux in Table~\ref{tab:observations} was calculated by integrating over the range in which there is significant signal, namely 15-35 \AA.
We briefly present here the {\sl Chandra} grating spectra already analyzed in \citet[][the Spectral Paper by members of this group]{Mitrani2025}, because their basic features and evolution are extremely useful for the timing analysis. However, these spectra are very complex and rich in interesting diagnostic atomic transitions, so the Spectral Paper presents the rigorous analysis, focusing on the underlying atomic physics.
\begin{figure}
\centering    
\includegraphics[width=0.8\columnwidth,height=2.0in]{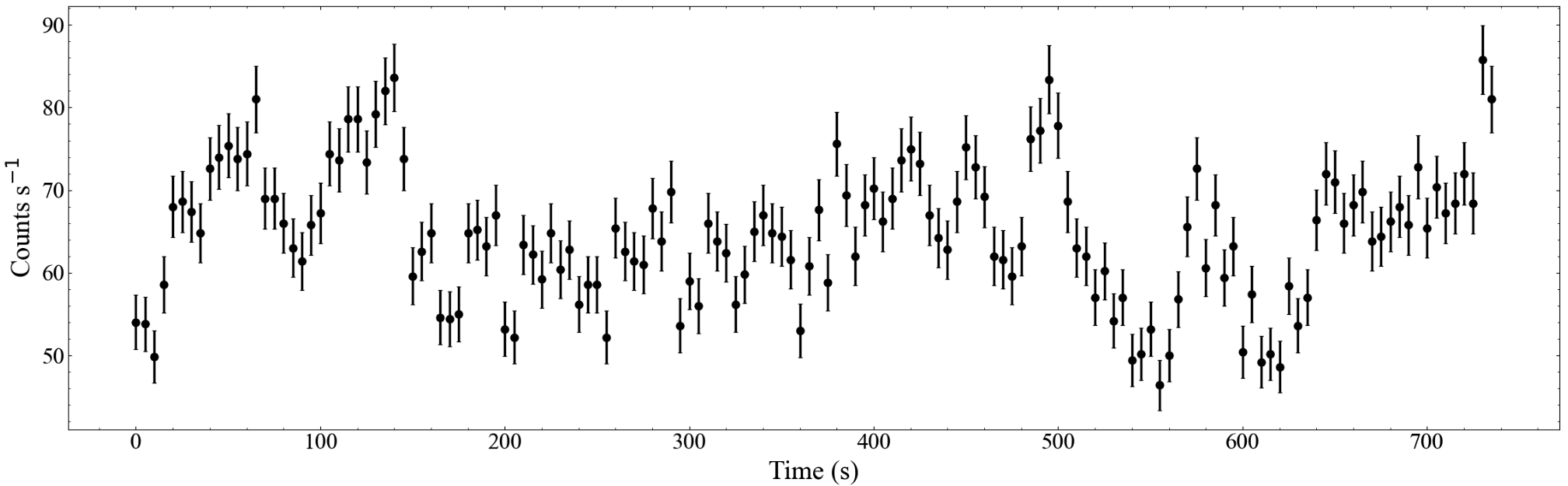}
\includegraphics[width=0.8\columnwidth,height=2.0in]{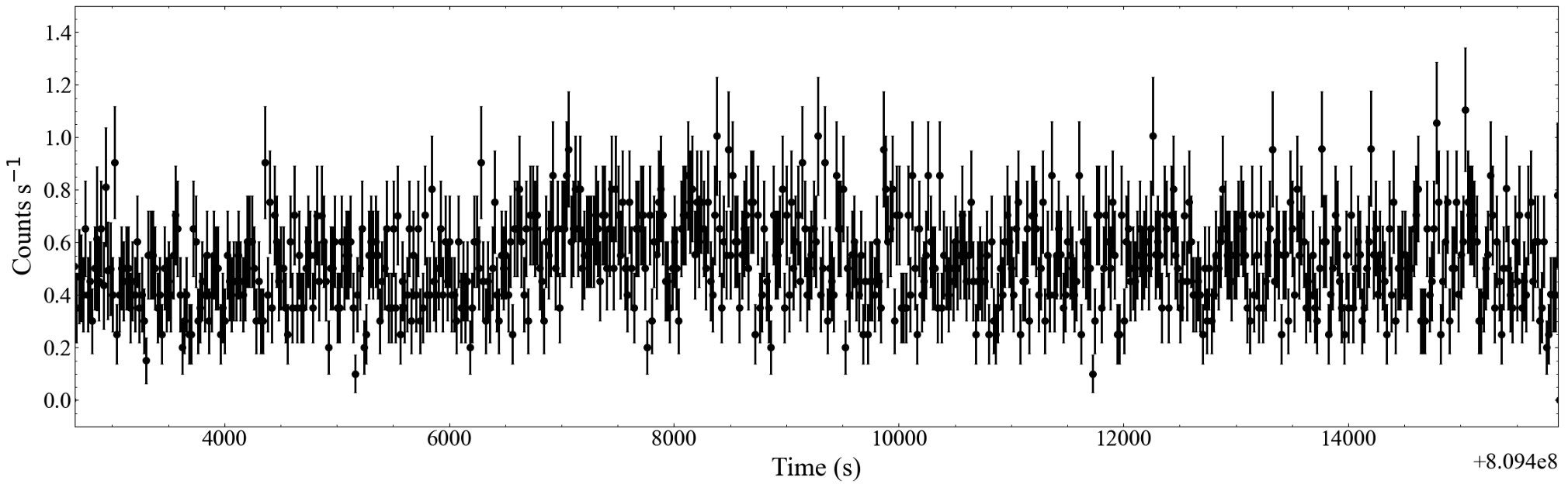}
\includegraphics[width=0.8\columnwidth,height=2.0in]{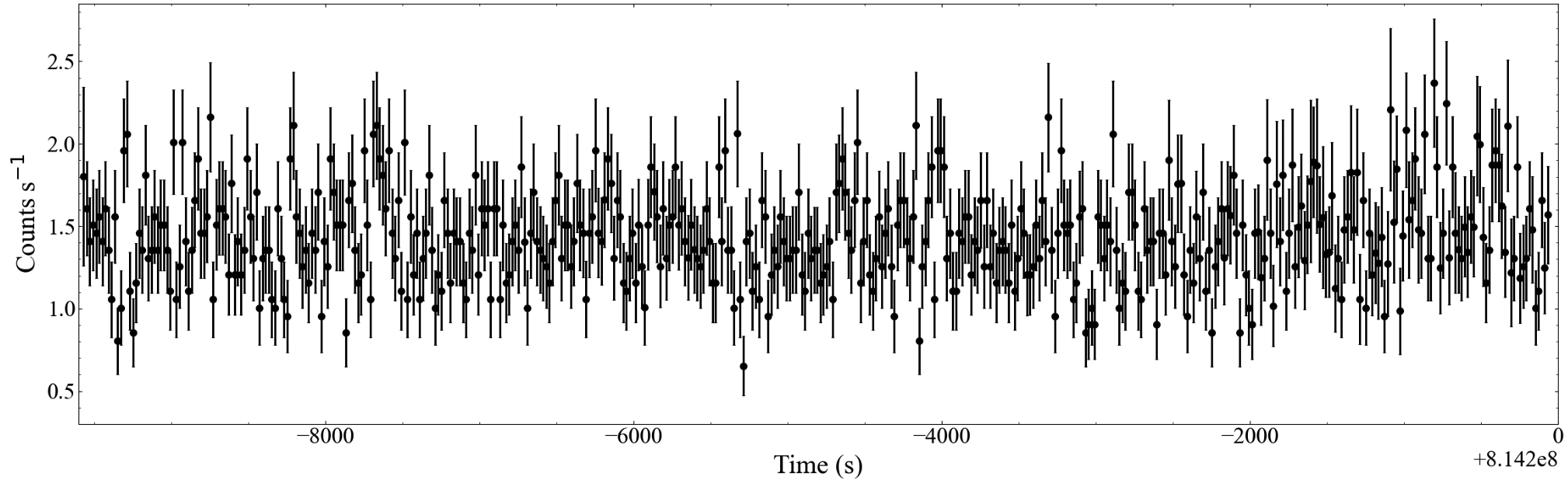}
\caption{Selected light curves to show the variability over time scales of minutes and of hours. The {\sl NICER} data from day 97 post-outburst (PO), in 5 second bins (top panel), and two {\sl Chandra}-HRC (zero order) light curves, respectively obs. 28048 (2023 August, middle panel, binned every 20 s) and 28049  (2023 October, lower panel, binned every 20 s). There is still some aperiodic variability in the {\sl NICER} data of the same day, but the average in each pulsation does not vary by more than 30\% while we observe fluctuations as large as  $>100$\%  over timescales of days after the first peak.}
\label{Stable_curves}
\end{figure}
\section{The large aperiodic variability}
Fig.~\ref{fig:light_curves} shows the global, long-term X-ray light curves, with large irregular variability. This variability occurs over timescales of days, while over timescales of hours there are fluctuations of smaller amplitude that were determined to be periodic \citep{Dethero2023, wang2024}. The lightcurve in the lower panel of Fig.~\ref{fig:light_curves} was presented in \citet{wang2024}, but the irregular variability was not discussed by these authors; in this paper we want to address both the irregular fluctuations and the short-term periodicity.
Fig.~\ref{Stable_curves} shows instead examples of the variability over timescales of hours: the average count rate is in most cases fairly stable in these short time intervals, while the periodic variability seems evident even by visual inspection.

\begin{figure}
  \centering
  \includegraphics[scale=0.3]{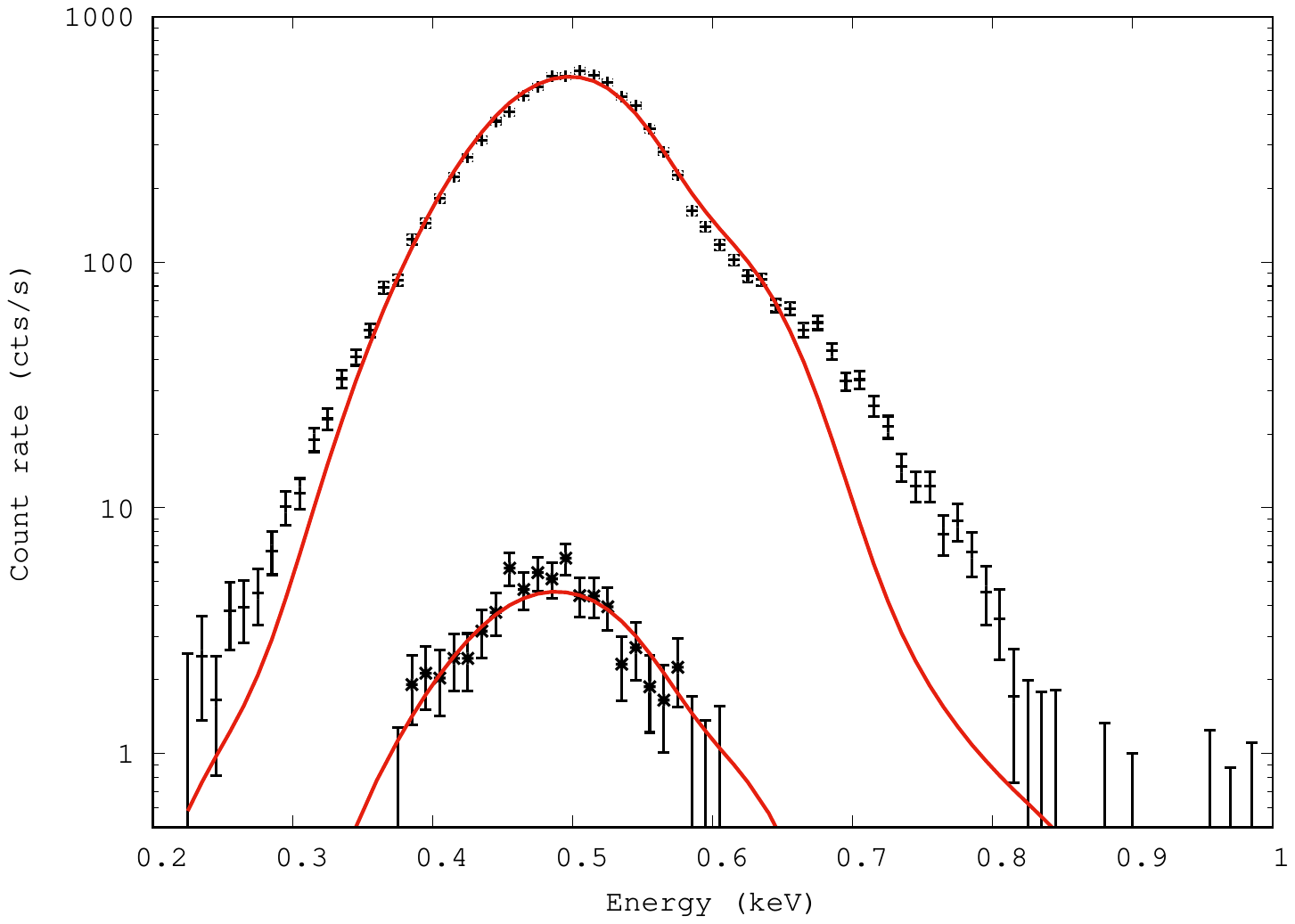}
  \includegraphics[scale=0.3]{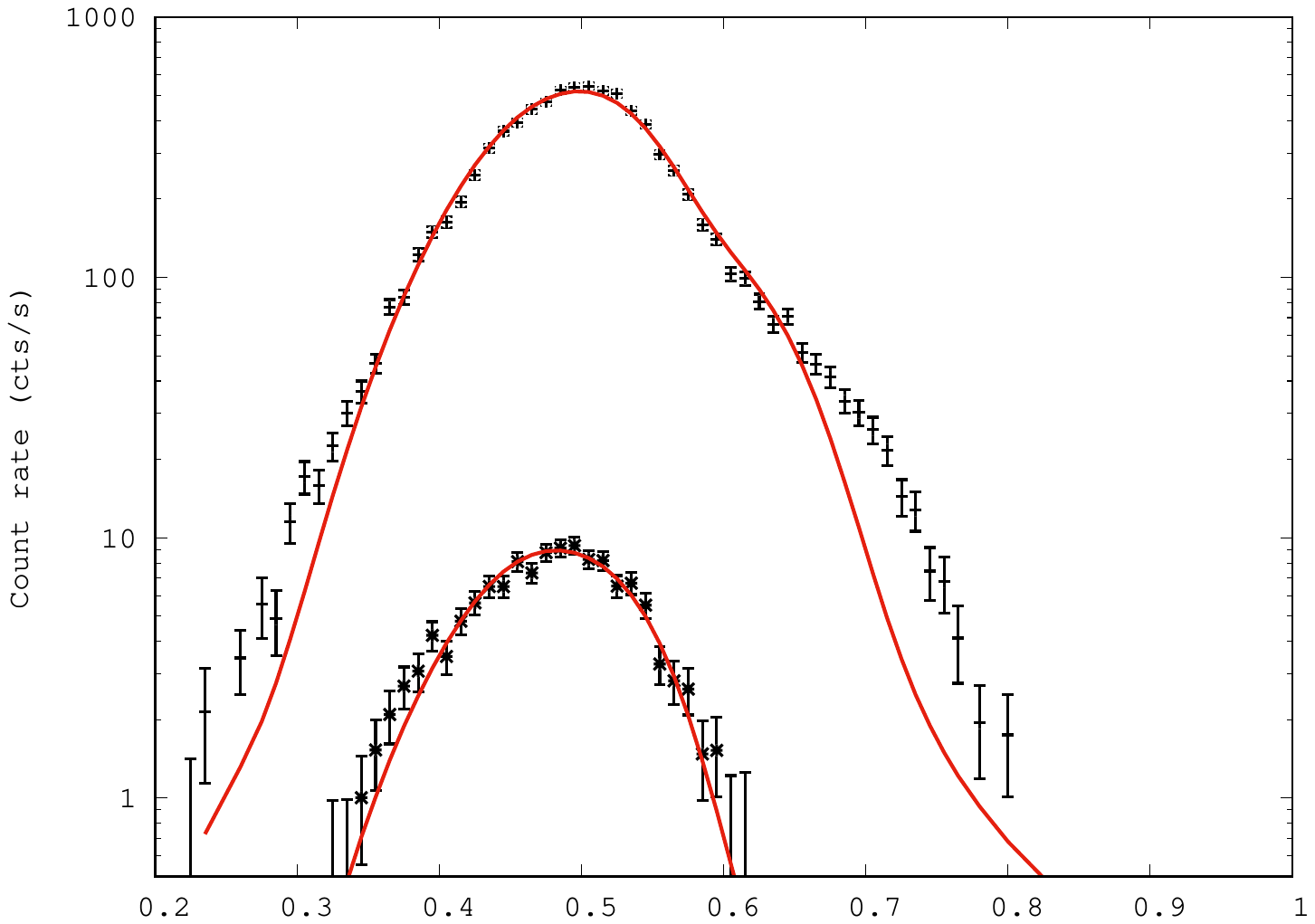}
  \caption{Comparison of high vs low count rate {\sl NICER} spectra: on the left, the spectrum measured on day 96 PO(high count rate) versus the one measured on day 102 (low count rate), and on the right, spectrum of day 118 (low count rate) compared to that of day 135 (high count rate; see Table~\ref{tab:NICEROBS}). In red, on the left, model fits with a temperature 
   of T=778,000 K and  N(H)=4.4 $\times 10^{21}$ cm$^{2}$ for the low count rate and N(H)=3.4 $\times 10^{21}$ cm$^{2}$  for the high count rate, respectively; on the right, a model fit with a fixed value 
   of the column density, N(H)=4.7 $\times 10^{21}$ cm$^{2}$,
    and T=782,000 K and 703,000 K for the high and low count rate, respectively.
  The logarithmic scale on the Y axis shows the deviation from
   a pure atmosphere model, that in the grating observations turned out to be due mostly to emission lines.}
  \label{fig:Comparison of NICER c}
\end{figure}

We know from theory and observations of several other novae, that in the post-nova SSS the WD temperature rises as the radius shrinks, then it is constant for a large portion of the SSS observability period; finally it decreases within days as the WD cools and nuclear burning presumably ceases \citep[see examples of observations and models, respectively, in][]{Balman1998,Yaron}.
Although we note an initial sharp rise in the {\sl Swift} light curve, the
 aperiodic variability of the SSS flux that followed over timescales
  of days  is different from a linear evolution of the SSS with a final decay due to cooling. However, it is not an unprecedented phenomenon in the lightcurves of novae-SSS
   \citep[e.g.,][]{Ness2003, Ness2011, Ness2023}.
Fig.~\ref{fig:Comparison of NICER c} shows some {\sl NICER} spectra with high and low count rate,
 and examples of fits with model 003 of the grid studied for Nova
  V4743 Sgr by \citet{2010ApJ...717..363R,rauch2010web}. It is clear that the atmosphere model alone does not fit the {\sl NICER} spectra,
   which is evident especially in the spectra taken with high count rate. Fig.~\ref{fig:Comparison of NICER c} show that,  using broad band spectra,
    it is not quite possible to differentiate between variations in column density, temperature and/or flux,  since with both hypotheses we reproduced the energy peak of the spectra.
      The {\sl Chandra} high resolution spectra shown and discussed in the next section highlight the complex physics of the observed spectrum with much more detail; we refer to the Spectral Paper for the analysis of the emission lines superimposed on the WD atmospheric continuum, resolved only in the high resolution grating spectra.
      
      Although the {\sl NICER} observations did not continue during
       the time between the two observing runs done with
 {\sl Chandra}, the {\sl Swift} monitoring went on: we see in Fig.~\ref{fig:light_curves} that,
 as the flux increased, the irregular variability decreased in amplitude during the month of 2023  October, before the last three {\sl Chandra} observations, although the
  observations were sparse (see Table~\ref{tab:Swiftdata}).
  In 2024 February the observations were resumed with {\sl Swift}; the XRT count rate was 3.563$\pm$0.067  cts s$^{-1}$
    on October 26 and  2.633$\pm$0.135 cts s$^{-1}$ on October 30,  but it had decreased to  0.320 $\pm$0.008  cts s$^{-1}$ when the observations were resumed. The XRT count rate kept on decreasing significantly in 2024 February, and after 2024 April there was no detection, as shown in Table~\ref{tab:Swiftdata}, so we inferred that the SSS turned off during this time  (see Section 6, Discussion).
 
 \section{The Chandra high resolution spectra}
\begin{figure}
  \centering
  \includegraphics[width=0.72\columnwidth,height=2in]{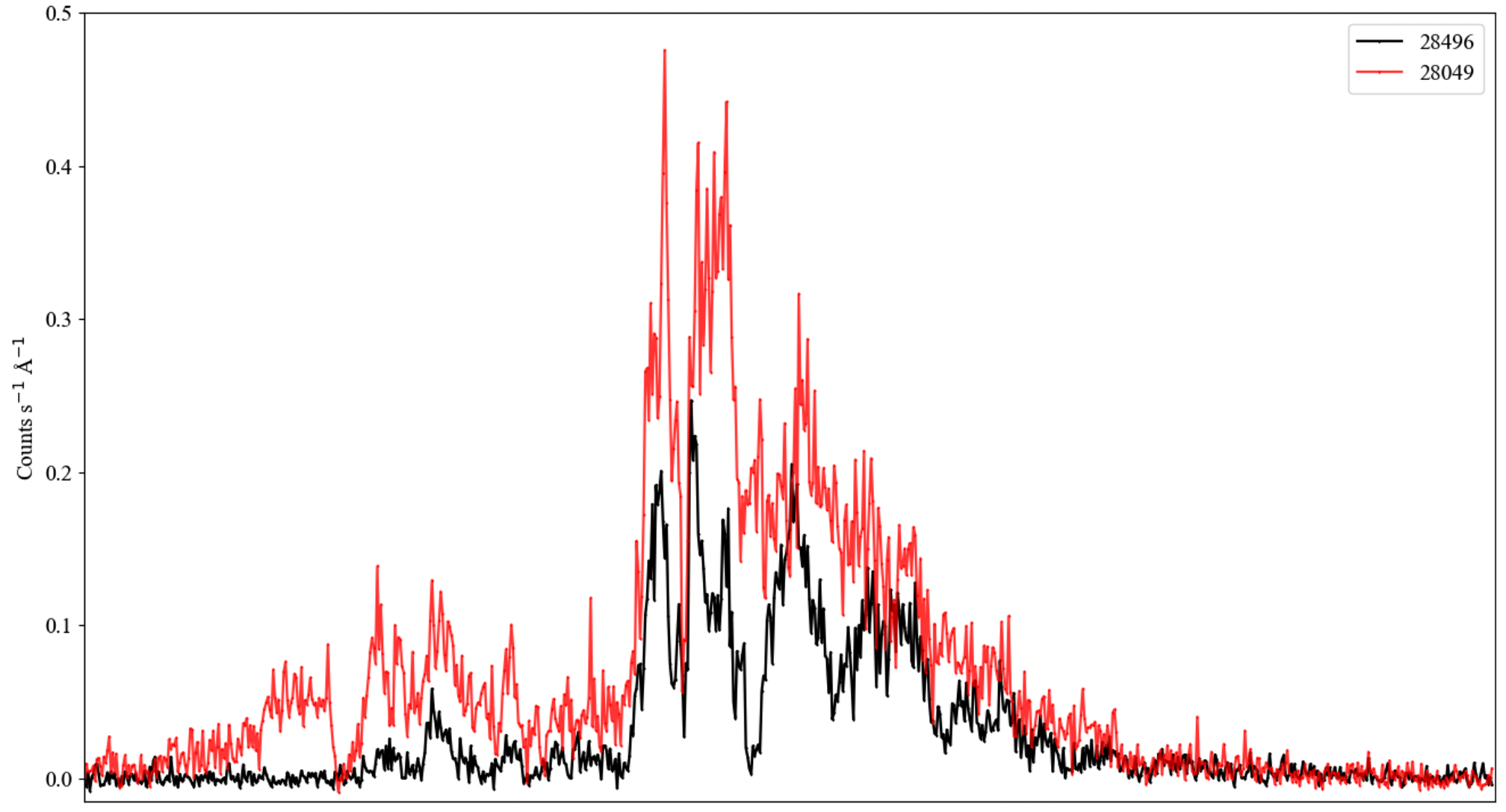}
  \includegraphics[width=0.72\columnwidth,height=2in]{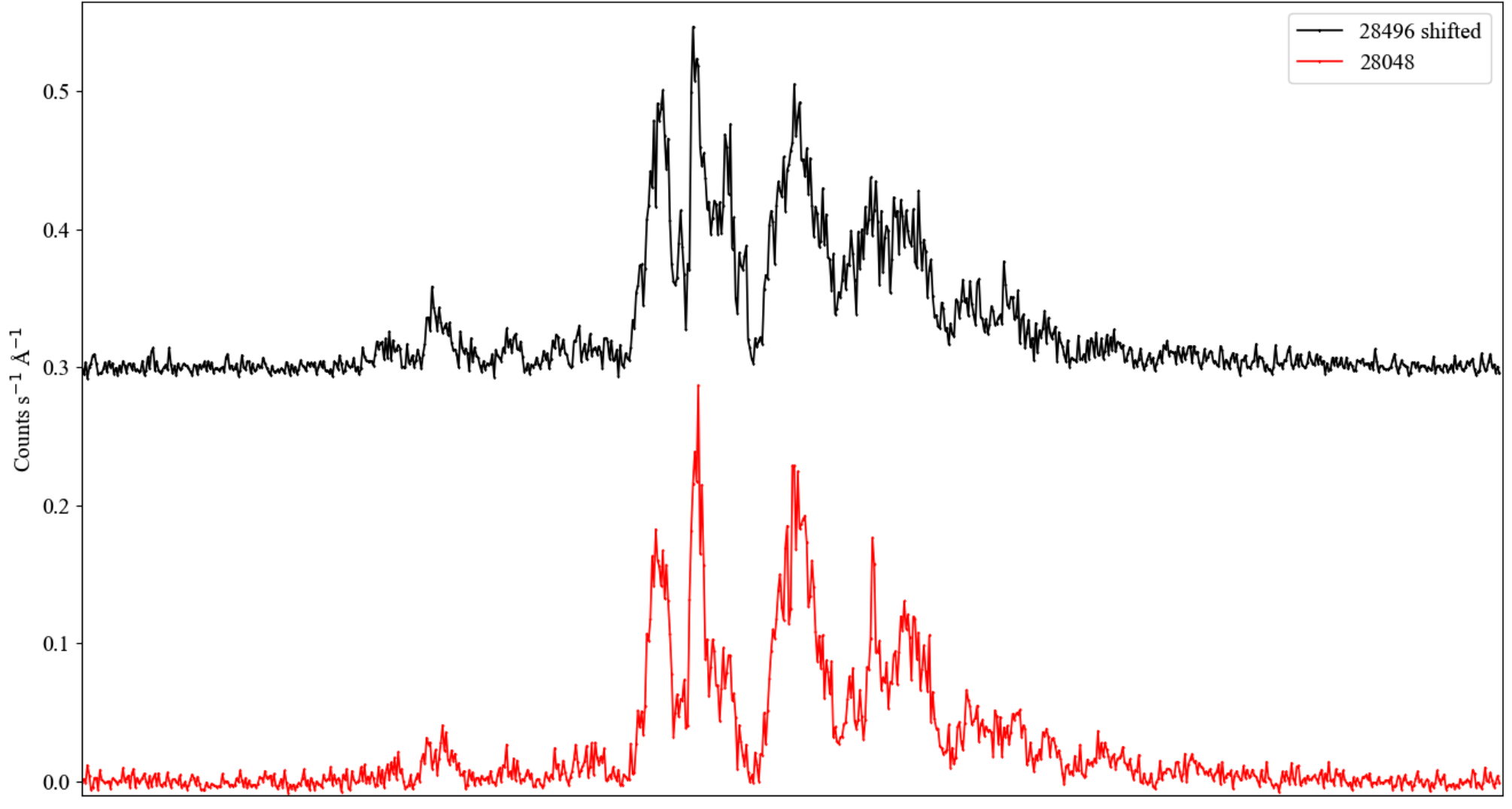}
  \includegraphics[width=0.74\columnwidth,height=2in]{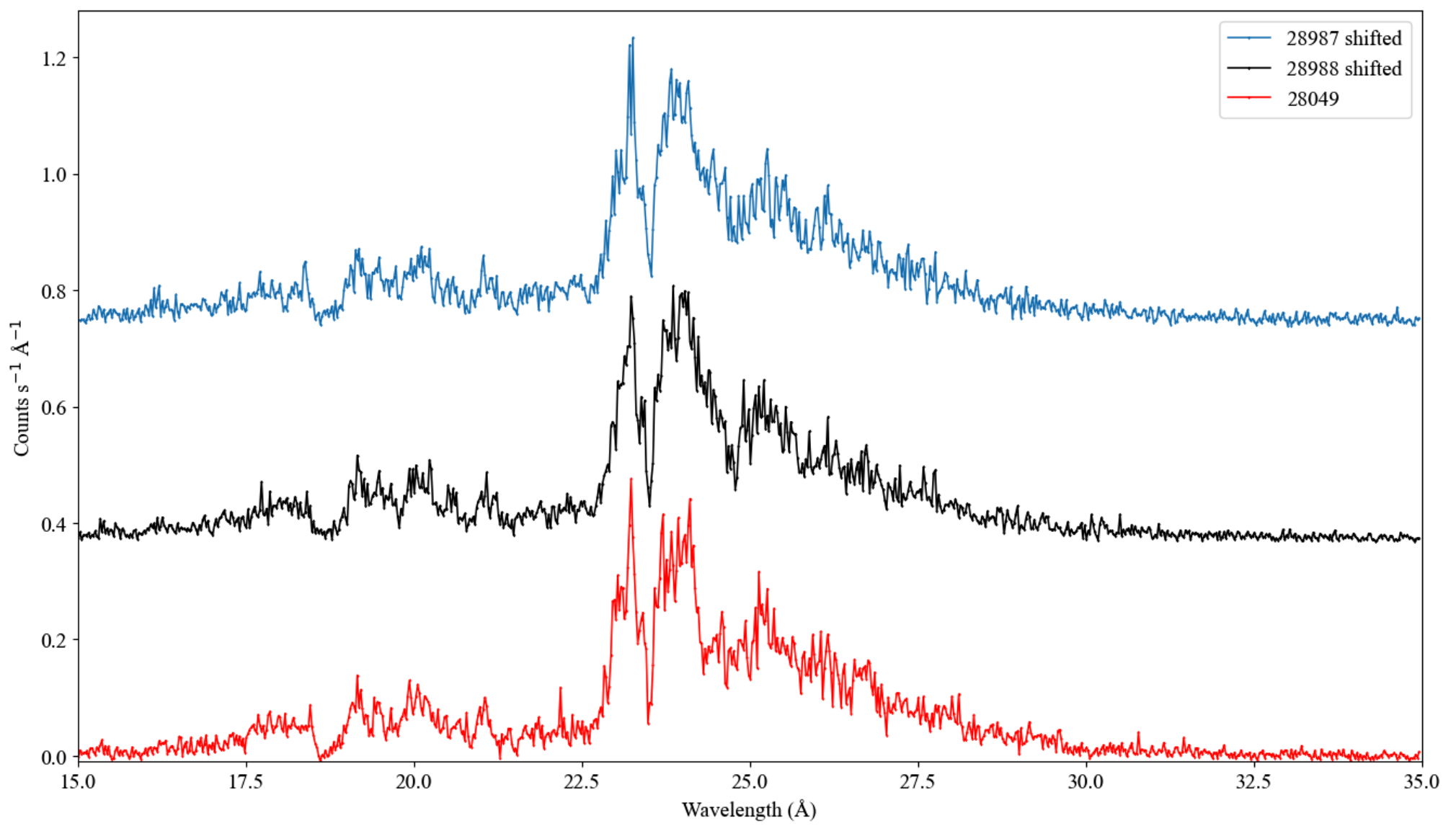}
\caption{Top panel: comparison of the $\pm{1}$ orders average spectra of August 27 (black) and October 20 (red). Middle panel: comparison of the August 26 (red) and August 27 (black) $\pm{1}$ orders spectra, with the second shifted by 0.375 cts s$^{-1}$. Bottom panel: comparison of the October 20 (red), 21(blue), 22 (black) $\pm{1}$ order spectra with successive shifts by 0.375 cts s$^{-1}$.
Note how the difference is significant only between August and October, but not in spectra taken on sequential days.}
\label{fig:Comparison of August to October Spectra}
\end{figure}

There was a factor $\sim$3 increase in measured flux in the October exposures compared with August; this is evident in two representative spectra of the two epochs displayed in the top panel of Fig.~\ref{fig:Comparison of August to October Spectra}. 
The change in the deep N VII absorption feature observed around 24.5 \AA\ in August, and not in October, resembles the spectral changes in RS Oph at different epochs after the outbursts (Fig. 6 of \cite{Ness2023}). In that paper, this type of variability was 
 attributed to variable ionization stages in ejecta with the same chemical composition. 
   In the Spectral Paper, we found that the large difference in flux was not due to the WD becoming hotter and
 more luminous in the time interval between the end of August and the end of October, indicating a completely opaque intervening
 absorber in August (see Fig.~\ref{fig:Comparison of NICER c}). We also note an increasing number of emission lines at shorter wavelengths in October. Such emission lines do not originate from the WD in any of the models with an effective temperature $>$200,000 K
  \citep[see][]{2010ApJ...717..363R, vanRossum2012}.
\begin{figure}
    \includegraphics[width=\columnwidth]{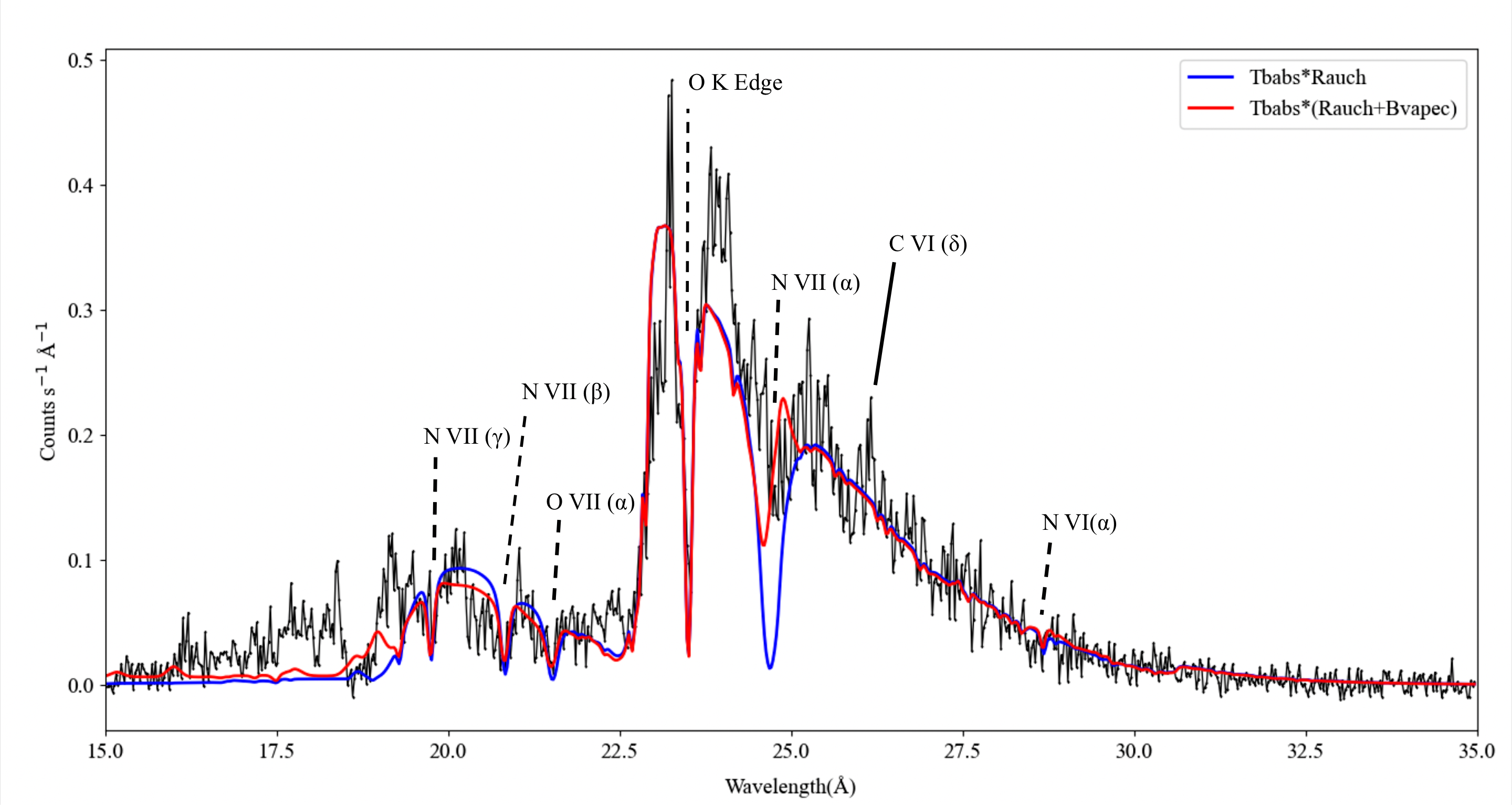}
    \caption{Fit to the spectrum taken on 2023 October 21 with the model described in the text. The blue line indicates the fit with only the atmosphere model. The BVAPEC model (included in the red line) adds emission lines of N VII and of oxygen, but it does not significantly improve the fit, as most emission lines are not fitted. A thorough analysis on the origin of emission lines - confirming the result on the atmospheric temperature - is presented in detail in the Spectral Paper.}
    \label{fig:28987model}
\end{figure} 
\subsection{The main source of X-rays: the white dwarf} \label{sub:SpecModel}
 In the Spectral Paper \citep{Mitrani2025}, we focused
  on a detailed analysis of the spectra and the atomic physics inferred from the emission lines. However,
   we need to outline and show that the main source of the continuum
   flux is the central SSS, namely the WD. In the Spectral Paper, in fact, we fitted the spectra with a composite physical model in XSPEC \citep{Arnaud}. The
2023 October spectra show a more luminous continuum and appear to have less relative contribution of flux in emission lines relative to the continuum. In Fig.~\ref{fig:28987model} we show the fit to one of these spectra with a Non-Local Thermal Equilibrium (NLTE) atmosphere model from the grid
calculated by \citet{2010ApJ...717..363R}  for Nova V4743 Sgr. We note that
the models are static and do not include an observed blue shift that we could clearly observe in
 the absorption features; however, an average blue shift velocity of $\approx$ 3500 km s$^{-1}$ is measured in the Spectral Paper in the 2023 August spectra by comparison with the \citet{2010ApJ...717..363R} model; a lower blue shift velocity of
 $\approx$1500 km s $^{-1}$ was inferred instead in the 2023 October spectra.
 
Fig.~\ref{fig:28987model} shows that from fitting the spectra as previously done for other novae with superimposed emission lines in the SSS phase,
\citep{Nelson2008, Orio2013, Tofflemire2013, Drake2016, Peretz2016, Orio2020, OrioBehar22}, namely by adding a thermal plasma in collisional ionization equilibrium (attributed to shocked ejecta in colliding winds) to the model for the luminous
 WD continuum, we could not reproduce the emission lines. For this reason we dedicated the Spectral Paper to the complex atomic physics producing the emission lines in these 
 {\sl Chandra} LETG spectra. In that work, we identify emission lines, including radiation recombination continua and charge exchange emission originating in the nova outflow interacting with cold gas in the circumstellar or interstellar medium. Here, we show in Fig.~\ref{fig:28987model}, how a model
 with logarithm of effective gravity $\log g=9$, specifically model S003 in the \citet{2010ApJ...717..363R}  (the blue line in Fig.~\ref{fig:28987model}) fits the general shape of the continuum. The residuals in the fit are due to the intricate pattern of emission lines discussed in the Spectral Paper.
The atmospheric model was multiplied by the TBABS model
 in XSPEC, to account for absorption due to the hydrogen column density along the line of sight \citep{2000ApJ...542..914W}.
 Because this is relevant for the interpretation of the
 timing analysis that follows, we note that the fits to the WD continua are obtained with effective temperature in the 750,000 K -- 780,0000 K range for each of the three October spectra, and
column density less than 20\% in excess of the interstellar value along the line of sight, which is $4.38\times 10^{21}$cm$^{-2}$
\citep{HI4PI2016}. The total absorbed flux ranges between 8.3 and $8.6 \times$ 10$^{-10}$ erg cm$^{-2}$ s$^{-1}$, about 90\%  of the measured value (see Table~\ref{tab:observations}). The value of the unabsorbed flux is 2.7-$2.8\times10^{-8}$ erg cm$^{-2}$ s$^{-1}$, and the emission lines contribute to $\le$4\%
 of the flux (see Spectral Paper for details).

\section{Timing Analysis: the periodic variability}
As mentioned above,
 \citet{wang2024} found quasi-periodic variability with an
 average period of $79.10 \pm 1.98$ seconds in the {\sl NICER} observations, in agreement with the previous analysis by \citep{Dethero2023}. The high signal-to-noise ratio of the {\sl NICER} spectra allows estimation of the modulation in all observations made with this instrument, but it is not always measurable in exposures with the lower count rate obtained with {\sl Swift} and {\sl Chandra}. However, we show below that the {\sl Chandra} lightcurves clarified the issue of the possible period drift, thanks to the long uninterrupted exposures.
 
 The analysis of \citet{wang2024}
 was based on the Lomb-Scargle periodogram method.
 These authors also listed the periods obtained for the single observations, and folding the short exposures' lightcurves with each obtained period seemed to indicate that the drift was real, implying that 
  a quasi-periodic oscillation (QPO) was measured, rather than a periodic modulation.
However,  an apparent drift can be an artifact of estimating a period of varying
 amplitude and short duration of individual light curves.
  \citet{dobrotka2010, dobrotka2017} studied the periodic variability of nova V4743\,Sgr - a nova
for which a complicated periodogram pattern was
 obtained in several observations in outburst -  in order to
determine whether the light curve was modulated with a single
signal or with multiple ones, and whether the periodicity
 evolved in time; in the second paper, 
 \citet{dobrotka2017} were able to demonstrate that
 the complicated periodogram patterns were the
 result of variable amplitude of a modulation with a constant period.
 
\begin{figure}
\resizebox{\hsize}{!}{\includegraphics[angle=0]{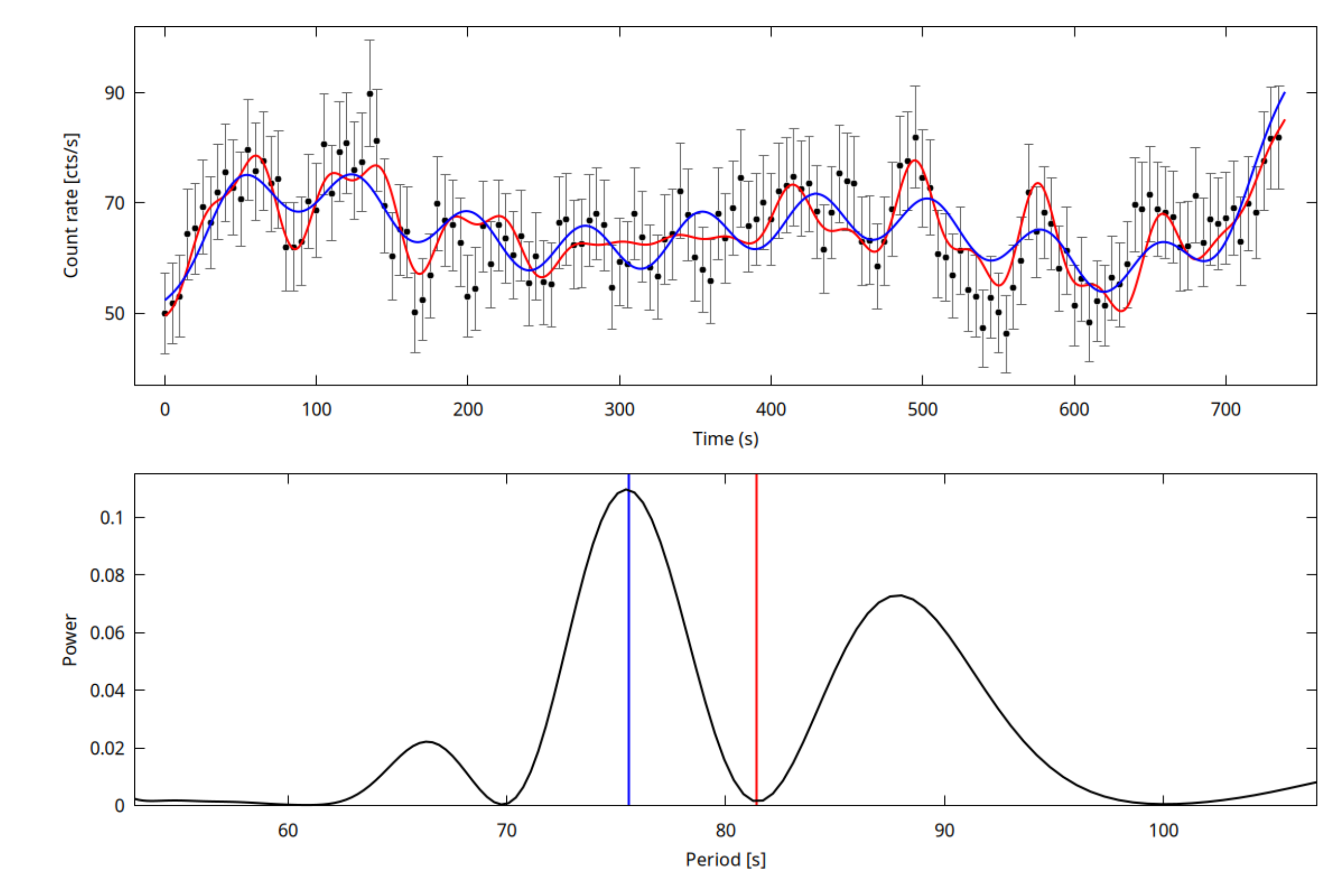}}
\caption{The black dots in the top panel show the light curve of {\sl NICER} observation 6203910103, binned every 5 s for better visualization.
The lower panel shows the corresponding Lomb-Scargle periodogram (black line). The colored lines represent the model fitting (in the upper panel) and derived corresponding periods  (in the lower panel);  the blue lines are for the single sine model with constant periodicity and constant amplitude (Equation 1), while the red lines show the result of the multi-component model with two sine functions, constant periodicity with its first harmonics, and variable amplitude (Equation~\ref{equation_fit_multicomp}).}
\label{lc_double_peak}
\end{figure}

 Since the Lomb-Scargle algorithm is based on simple sine waves
 (\citealt{scargle1982}, \citealt{vanderplas2018}),
 the resulting periodogram must be taken with caution. We take as an example the {\sl NICER} observation of V1716 Sco
 number 6203910103 (see Table~\ref{tab:NICEROBS}), for which
 the Lomb-Scargle periodogram yields a period of 75.69 s
 according to the analysis of \citealt{wang2024}.
The light curve of this
 observation, with the corresponding periodogram, is 
 in Fig.~\ref{lc_double_peak}. Our
 analysis shows a clear double peak,
 with the same type of pattern studied by \citet{dobrotka2017}.
 As shown by these authors, when the amplitude of the modulation varies, the double peak pattern is often an artifact
 of the amplitude variation, while the frequency is instead unique and constant,
  and the actual value is between the two peaks resulting from the Lomb-Scargle analysis. For this specific exposure, \citet{wang2024}
 listed in their table the period corresponding to the highest peak.
 We experimented with a different method from the Lomb-Scargle by fitting different functions to the light curves. We first tried
 assuming a (constant) period $p$ and (constant) amplitude $a$
 as free parameters, and a sinusoidal curve and a
fifth-order polynomial
 P$_{\rm t}$($t$); 
\begin{equation}
\psi_{S} = a\,{\rm sin}(2\,\pi\,t/p + \phi) + {\rm P}_{\rm t}(t),
\label{equation_fit_singlecomp}
\end{equation}
 where $\psi$ is the count rate value and the units of time are in seconds. This is the blue line in the upper panel of Fig.~\ref{lc_double_peak}. This fitting method yields the same period given by \citet{wang2024}.
  However, in this observation, we see that after the first 200 s the amplitude
 of the observed modulation varied; it was smaller between approximately 200 and 400\,s into the exposure. This suggests that fitting with a model of a sine function with constant amplitude is not quite adequate. Thus, in order to test the hypothesis of variable amplitude, we fitted the light curve with a more complex model,
 in which the light curve is described by a fifth-order polynomial\footnote{It is the same as in Equation~(\ref{equation_fit_singlecomp}). We tried also different orders and the results are robust, i.e. the derived periodicities are within the errors.}
 P$_{\rm t}$($t$), the variable amplitude of the sine function
 is fitted with another polynomial P$_{\rm a}$($t$) of
 the fifth order, and an additional
 component with first harmonics, with a fraction $f$ of the
 fundamental's amplitude is added:
\begin{equation}
\psi_{M} = {\rm P}_{\rm a}(t)\,\left[ {\rm sin}(2\,\pi\,t/p + \phi_1) + f\,{\rm sin}(2\,\pi\,t\,2/p + \phi_2) \right] + {\rm P}_{\rm t}(t).
\label{equation_fit_multicomp}
\end{equation}
The fit \footnote{ We note that $\mathrm{e}{-x}$ here denotes $10^{-x}$ for brevity's sake. $ \psi_{S} = 4.30 \, \sin\Bigl(2 \pi t / 75.76 + 3.72 \Bigr) + (54.68 + 0.528\,t - 5.09\mathrm{e}{-3}\,t^{2} + 1.86\mathrm{e}{-5}\,t^{3} - 2.88\mathrm{e}{-8}\,t^{4} + 1.59\mathrm{e}{-11}\,t^{5}) \newline \psi_{M} = (3.86 + 8.18\mathrm{e}{-2}\,t - 5.39\mathrm{e}{-4}\,t^{2} + 1.07\mathrm{e}{-6}\,t^{3} - 1.12\mathrm{e}{-9}\,t^{4} + 6.06\mathrm{e}{-13}\,t^{5}) \cdot [\sin(2 \pi t / 81.79 + 4.21) + 0.51 \, \sin(2 \pi \cdot 2 t / 81.79 + 4.30)] + (54.68 + 0.528\,t - 5.09\mathrm{e}{-3}\,t^{2} + 1.86\mathrm{e}{-5}\,t^{3} - 2.88\mathrm{e}{-8} \,t^{4} + 1.59\mathrm{e}{-11}\,t^{5}
)$} of this model is shown as the red line in the upper panel of Fig.~\ref{lc_double_peak}. This function, unlike the previous one, describes the reduced amplitude between 200 s and 400 s. \textbf{We attribute the improved fit obtained with this model primarily to the inclusion of P$_{\rm a}$($t$), and note that the component with first harmonics only marginally further improves the result.} In this fit, the resulting periodicity is 81.41\,s, which falls
 exactly in the middle between the two periodogram peaks shown in the
 lower panel of Fig.~\ref{lc_double_peak}. For each $\text{exposure} \geq \text{500}$ s, the goodness of the fit of both models (simple and multi-component) as $\chi^2_{\text{red}}$ was calculated. Additionally, the Akaike Information Criterion (AIC) was calculated for both models ($AIC_S$ for simple and $AIC_M$ for multi-component), supporting the multi-component model as the preferred one. The $\chi^2_{\text{red}}$ values together with the $\Delta AIC = AIC_S - AIC_M$ are included in Table \ref{tab:NICEROBS}.
 
\begin{figure}
\resizebox{\hsize}{!}{\includegraphics[angle=0]{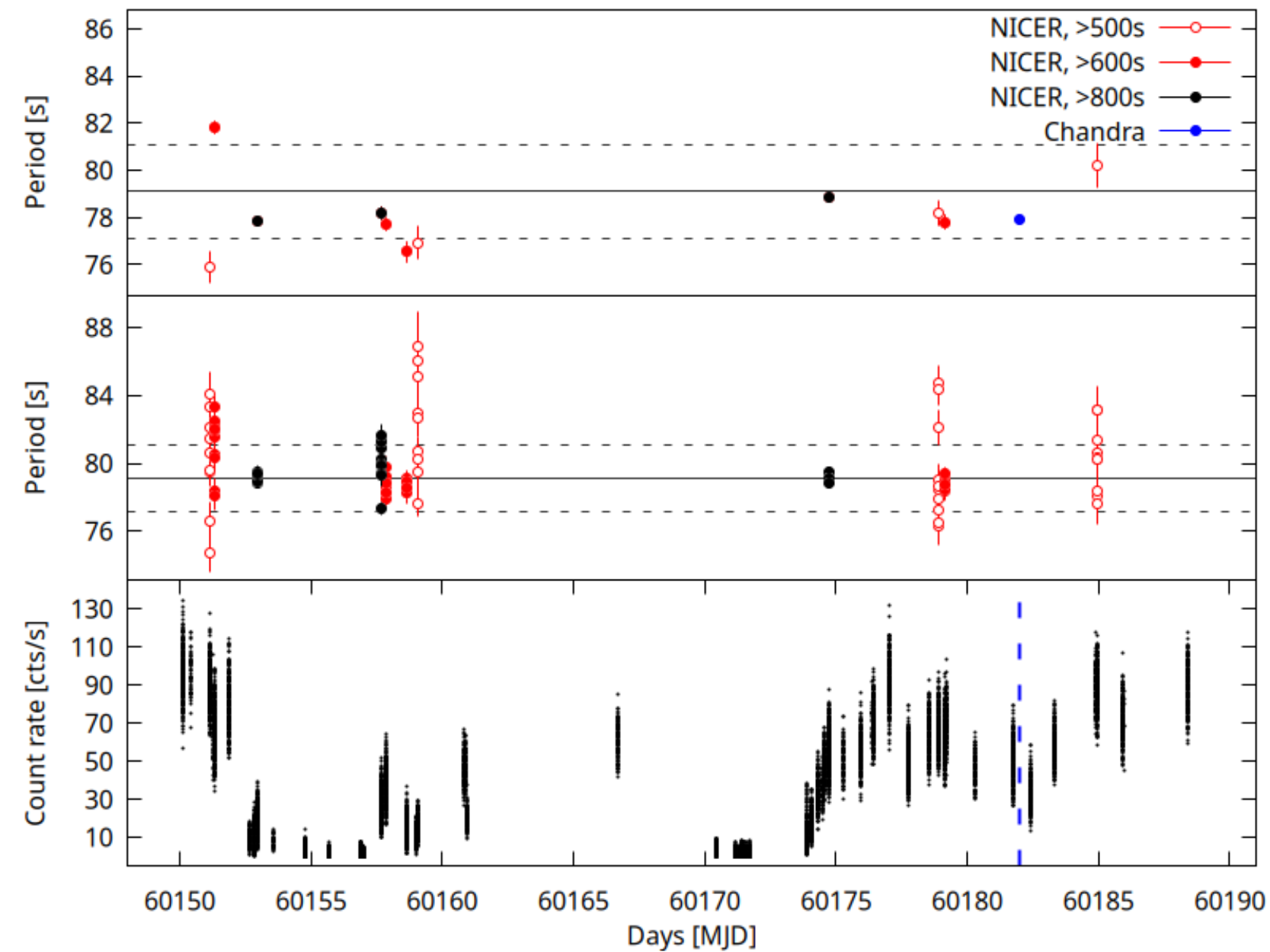}}
\caption{Upper panel - Measured periods with multi-component fitting using Equation~(\ref{equation_fit_multicomp}). The errors were evaluated  using the Python Numpy package \citep{harris2020}. The style and colors of the points represent the different ranges of exposure times. The horizontal lines indicate the mean value and error interval derived by \citet{wang2024}. Middle panel - the results (with the colors adopted in the upper panel) using simulated light curves in which the period is constant, to show the resulting scatter of the measured period in the simulation. The scatter is larger for the shorter exposure times.  Finally, the lower panel shows the whole, unbinned light curve for reference (In Fig.~\ref{fig:light_curves} we only showed the average count rate of each short exposure).}
\label{period_evolution}
\end{figure}

\begin{figure}
\resizebox{\hsize}{!}{\includegraphics[angle=0]{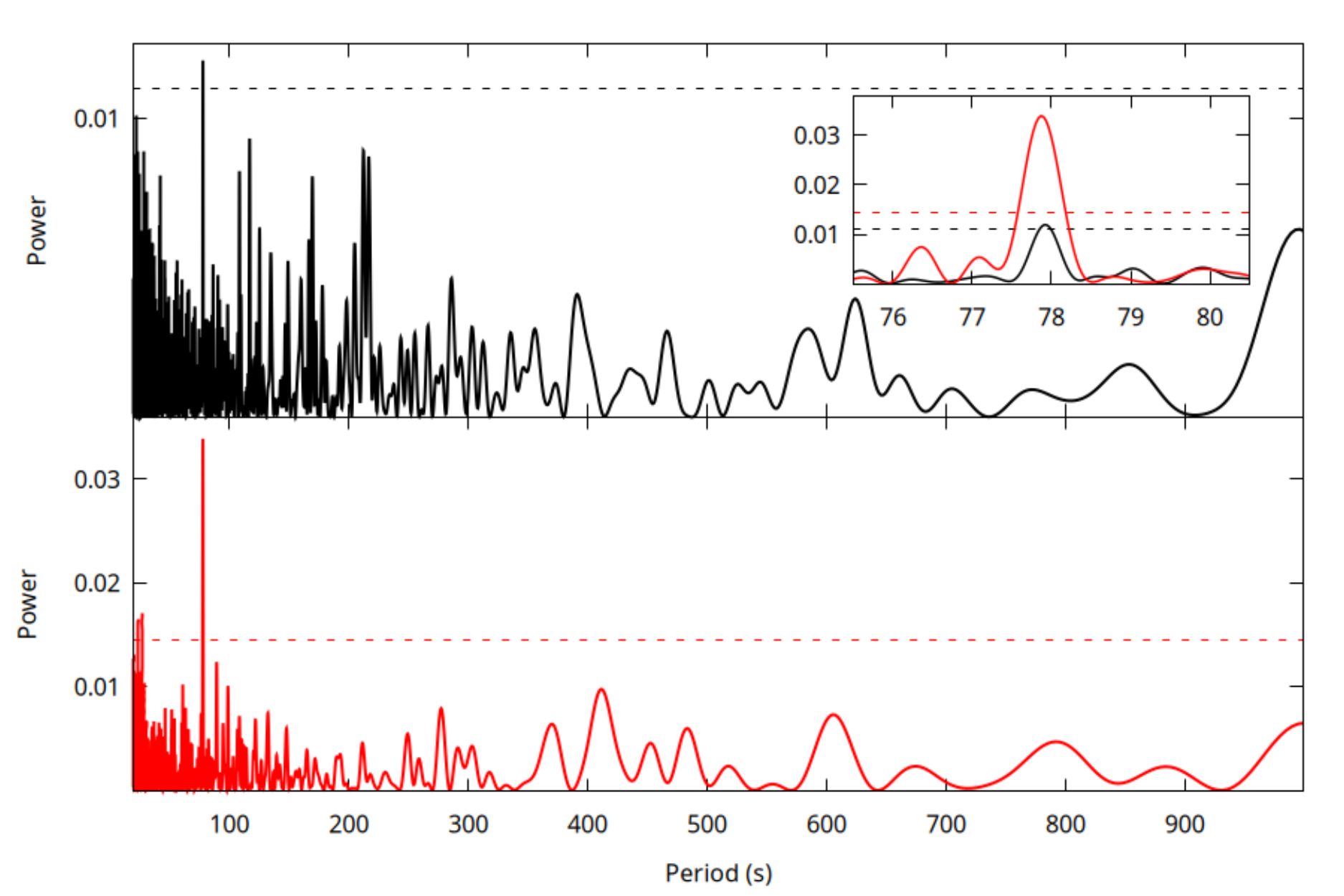}}
\caption{Periodograms of two {\sl Chandra} light curves. ObsID 28048 in the upper panel in black represents the observations during the {\sl NICER} interval in Fig.~\ref{period_evolution}, ObsID 28049 is shown in the bottom panel in red. The horizontal dashed lines mark the corresponding 99.9\% confidence level. The inset in the upper panel is the zoom of the two observations, showing the comparison of detected period around the 77.9 s.}
\label{period_chandra}
\end{figure}

Two {\sl Chandra} observations were done in August
 overlapping with the interval of the {\sl NICER} observations (see Tables A2 and A3). Only observation 28048 (August) yielded a detection above the 99.9\% confidence level \footnote{Since the false alarm probability (FAP) is dependent on the period interval, we computed the FAP from minimal periodicity corresponding to the Nyquist frequency up to 1000\,s.} around the studied period by \citet{wang2024} with the Lomb-Scargle method (which we indicate with a blue point in the top panel and with a blue dashed line in the bottom panel of Fig.~\ref{period_evolution}). The {\sl Chandra} HRC collects considerably fewer photons than {\sl NICER} and when the 
 amplitude of the modulation is small, we cannot detect the periodicity with high probability, so the non-detections do not imply that the modulation actually disappeared. Another {\sl Chandra} observation yielding a detection above 99.9\% confidence level is observation 28049 from 2023 October 20, while in the other two exposures of October the amplitude of the modulation was too small to obtain significant results. The periodograms for the two {\sl Chandra} observations yielding  99.96\% (ObsID: 28048) and 99.99\% (ObsID: 28049) detection level of the period are shown in Fig.~\ref{period_chandra}.
 Fitting Gaussian profiles to these two peaks, we also estimated
 the errors for the two {\sl Chandra} dates. The two periods we derive are
 $77.921 \pm 0.175$ s (August) and $77.875s \pm 0.231$ s
 (October).
 Since the {\sl Chandra} results are based on much longer continuous exposures than the {\sl NICER} data, we
 examined the hypothesis that the
 results of the long, uninterrupted exposures are indeed more reliable, and that the period drift derived from the {\sl NICER} data is an artifact of the varying amplitude of the modulation during the shorter exposures. 

\begin{figure}
    \includegraphics[width=\columnwidth,height=4.5in]{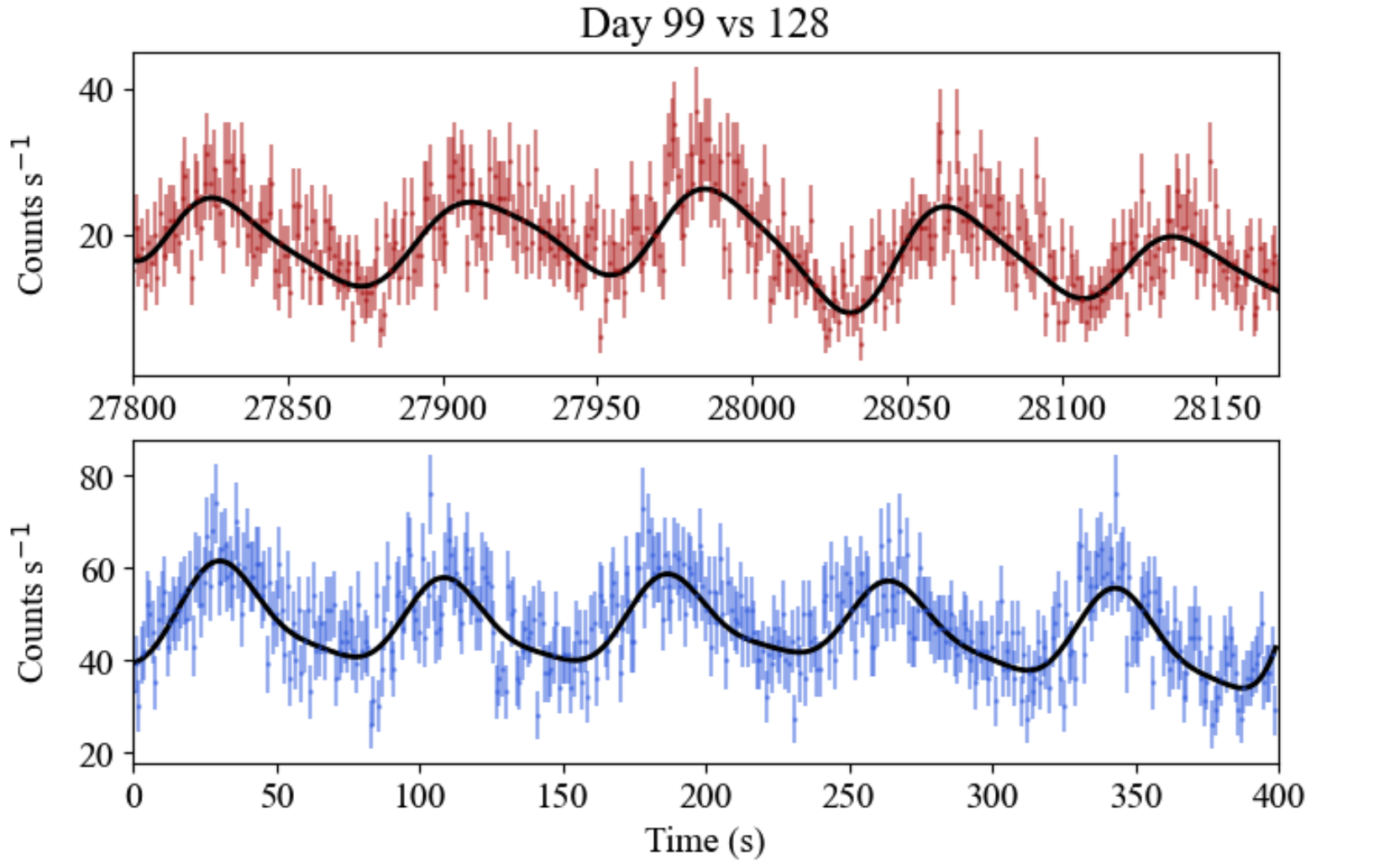}
    \caption{Light curve comparison of Day 99 (red) vs Day 128 PO (blue). In this segment of the Day 99 exposure, the average count rate is $\sim$17 ct s$^{-1}$, and the average count rate for Day 128 is $\sim$ 47 ct s$^{-1}$. We fit Equation \ref{equation_fit_multicomp} to both light curve segments (black lines) to estimate the modulation amplitude for each. We note that the fitting here is intended only to estimate the amplitudes, and not used in our timing analysis, since both segments are $\sim$ 400 s, below our 500 s threshold. The modulation amplitude for Day 99 is 34$\pm{5}$ percent of the average count rate, vs a modulation amplitude of 18$\pm{4}$ percent of the average count rate for Day 128. Various {\sl NICER} light curves with largely different average count rates show this dis-correlation when compared, and we suggest that the amplitude of the modulation on the periodic signal is not correlated with variations in observed flux from the source.}
    \label{fig:day99v128}
\end{figure}

 In previous work, the fact that the variability cannot always be measured with {\sl Chandra} and {\sl XMM-Newton} in the SSS phase of novae has been attributed to a ``disappearing modulation'' \citep{Ness2015}. However, the observations with high signal-to-noise obtained with {\sl NICER} have changed this perception in the case of RS Oph \citep[see][]{Orio2023}. Before we continue, we show in Fig.~\ref{fig:day99v128}  that the variation of the amplitude measured with {\sl NICER} is not flux-dependent. Therefore, even if the {\sl Chandra} count rate did not vary after 1-2 days' interval, it is reasonable to infer that the modulation amplitude was too small for a measurement, not that the fluctuation ``disappeared''. This will be further discussed in Section 6.
   
 In order to test whether a period with a modulation of
 variable amplitude in the short {\sl NICER} exposures 
can be recovered with a simple periodogram,
we experimented with fitting the {\sl NICER} light curves using the multi-component model in Eq.~(\ref{equation_fit_multicomp}). At first, we selected
  the {\sl NICER} observations in which the period was
   detected 
 with a confidence level above 99\%, and with an exposure of at 
 least approximately 10 cycles, i.e. 800 s.
 The resulting periods are shown in the upper panel of Fig.~\ref{period_evolution} (black points).
In the figure, we also added the results of the
 exposures with
 a minimum exposure time of 600 and 500\,s. The results are presented in different colors,
 depending on the lower limit of the exposure time in the observations, also in the upper panel of Fig.~\ref{period_evolution}. The figure shows a scatter around the estimated period around the value of 79.1 s
 measured by \citet{wang2024},
indicated by the solid horizontal line in the figure
 (with the dashed lines representing their error interval).
 The majority of our measurements are within,
 or close to, the error interval of \citet{wang2024}. 
%
%

Finally, we performed a simulation to examine whether the scatter in the results is only an artifact of the variable amplitude in the short exposures. Starting from our multi-component fits (Equation~\ref{equation_fit_multicomp}) to the {\sl NICER} observations assuming variable amplitude and higher harmonics, we
randomly modified\footnote{We added a random small value to all coefficients except the one determining the mean flux.} the coefficients of the polynomial P$_{\rm a}$($t$). The period was assumed to have a constant value of 79.1\,s, as determined by \citet{wang2024}. We simulated 10 light curves for each {\sl NICER} observation, i.e. we used the exposure time and time bins of each observation,  and only added Poisson noise. We selected only those simulated light curves where the maximum count rate was no higher than the maximum observed value, and vice versa for the minimum count rate, in order to avoid unusual trends, significantly differing from the observed light curve. We thus obtained a sample of  10 light curves of simulated {\sl NICER} data with varying pulsation amplitude,  but with a constant period, and performed the same fitting analysis  done for the observed data.  The middle panel of  Fig.~\ref{period_evolution} shows indeed very scattered results, that quite well match the scattering obtained by fitting the real {\sl NICER} data  (plotted in the upper panel of the figure). The duration of the exposure is very important,  because
the most scattered results are those corresponding to the shortest exposures, while the less scattered ones are derived for the longer exposure times of more than 800 s. Clearly, the duration of the observation plays a decisive role in this scatter, and the variability of the periodicity derived from the {\sl NICER} observations is probably an artifact of the short exposures and modulations of variable amplitude. Therefore, we rely on the much longer and continuous {\sl Chandra} exposures, and we conclude that the period was most likely stable, while the {\sl NICER} short duration of uninterrupted exposures makes these data unsuitable for the analysis of the stability of the period. 
\section{Discussion}
The SSS flux of Nova V1716 Sco was variable, both aperiodically and periodically.
Between late July and September 2023 there was large
 irregular variability over timescales of days. This aperiodic variability persisted but decreased in amplitude in the following month of October. The SSS flux detected
with {\sl Chandra} in the 15-35 \AA\ range increased approximately threefold between exposures at the end of August and at the end of October. 
   However, the energy of the peak of the flux and a preliminary fit with a WD atmosphere - supported by more detailed calculations including the emission lines in the Spectral Paper - indicate that the WD did not vary significantly in effective temperature between August and October; we suggest that in August the SSS flux was lower because  it was probably still partially obscured by clumpy ejecta, or other circumstellar material in the system. 
   
 The October {\sl Chandra} spectra fitted with a multi-component model in XSPEC utilizing the TMAP atmosphere grids 
indicate a WD temperature of approximately 750,000-780,000K, consistent with a massive white dwarf 
 according to the nova models by \citet{Starrfield2012, 2013ApJ...777..136W}. The detailed spectral analysis is contained in the Spectral Paper, which
 also found that the comparison with the atmosphere model indicates an average blueshift of the absorption features of the WD by
   $\approx$3500 km s$^{-1}$ in the August spectrum and by about
   $\approx$1500 km s $^{-1}$ in October.  We interpreted this as the signature of an expanding WD atmosphere. While radiation driven wind models like \citet{Hachisu2007} predict that mass loss ends at the beginning of the SSS stage \citep[around day 105 in this case, see ][]{2023ATel16069....1P}, blueshifted absorption lines in the atmospheres of the WD indicate that the atmosphere is still emitting a wind, even after the radius of the WD has significantly shrunk back to almost pre-outburst size
 \citep[e.g.][]{Nelson2008, Orio2016}.

  We note that the SSS irregular aperiodic variability has been observed in other novae. For instance, in V4743 Sgr the SSS disappeared completely at the end of an exposure and was observed again at the same flux level when the observations were restarted
  \citep{Ness2003};
 irregular variability observed in nova U Sco in its 2010 outburst was attributed to
 clumps in the ejecta that were completely optically thick to X-rays \citep{Ness2012}. 

Another possibility is that,
 if the system has high inclination, the accretion disk may not have been completely disrupted or it was reforming, 
  causing the variable absorption, but 
 we do not have dense enough data to measure a periodic modulation over timescales of days, that may occur over the orbital period. However, the shape of the light curve in Fig.~\ref{fig:light_curves} can hardly be reconciled with the scenario of a long period modulation of the flux, since exposures with very different intervals of time elapsing between them, often yielded about the same count rate, as can be seen in Table~\ref{tab:NICEROBS}
   reporting the {\sl NICER} average count rates.  

The timing analysis of the periodic modulations yielded a period of $\simeq$77.9 seconds, which can be measured in almost all
 {\sl NICER} exposures, but the amplitude is at times too small to be measurable with {\sl Swift} and {\sl Chandra}. We also find that the amplitude of the modulation in {\sl NICER} is not correlated with the average flux during a given exposure: the long-term irregular and the short-term regular variability appear to be completely uncorrelated. A possibility we suggest is that the modulation is due to variable visibility of an area that emits more flux, possibly at the polar caps.
 
 However, on days during which the amplitude of the oscillation was large enough to be clearly
 measurable with the {\sl Chandra} HRC detector, we found a constant period not only during the same day, but also
in exposures taken two months apart. We found that the Lomb-Scargle periodogram method can produce double peaks in periodograms when amplitude modulation is present, as demonstrated in Figure~\ref{period_evolution}. This result aligns with \citet{dobrotka2017}, who similarly demonstrated that amplitude variability in X-ray light curves of novae can create false signals or double peaks in periodograms. Using simulated light curves with added Poisson noise and duration of exposures similar to the actual  {\sl NICER} ones, we confirmed that the observed period is likely to be stable, with no significant drift across observations. Instead, the apparent period drift in the {\sl NICER} observations arises as an artifact of short exposures and variable modulation amplitude. To model the signal accurately, we applied a sinusoidal fit with a fifth-order polynomial and harmonics, demonstrating that the quasi-periodicity is consistent with a single signal.

   \citet{wang2024} folded the lightcurve of single exposures with their best period of 79.10 s,
   finding that a single period does not fit these lightcurves over time. However, we must consider that the {\sl NICER} data 
   span a period of approximately 38.25 days or $\sim$ 42,400 cycles. The uncertainty estimated for the period derived by these authors is large at $\pm{1.98}$ s, so it is understandable that the lightcurves of the individual observations are out of phase after so many cycles, and this does not sufficiently prove that the period actually varied.

A single
 signal, with a period of several minutes,  in several other novae with SSS modulations with longer periods has been interpreted as  the rotation period of the WD \citep{Leibowitz2006, Aydi2018, Drake2021}. This rotation-based modulation is consistent with an intermediate polar (IP), highly magnetized  WD (IPs usually have magnetic fields in the 10$^{5}$--$10^{6}$ Gauss range).  In fact, several novae with periodic modulations in the SSS phase were later found to have signatures of the
 magnetic accretion of IPs while accreting in quiescence, years after the outburst, or even before \citep{Zemko2015, Zemko2016, Zemko2018, Drake2021, Orio2024}. However, we still do not understand the exact mechanism for which the
magnetic fields shape the outflowing material and why
 the SSS flux is modulated with the WD spin period; we only note that bipolar outflows from the WD are likely to happen \citep{Chomiuk21}.
\section{Conclusions}
The main conclusions that can be drawn from our analysis of the X-ray data of Nova V1716 Sco are the following:

$\bullet$ The SSS phase started a little after 3 months since the onset of the optical eruption and lasted between 3 and 6 additional months. When {\sl Swift} monitoring resumed in early 2024 February, the count rate decreased by an order of magnitude and moderate cooling with respect to late October is estimated with spectral fits. This was followed by measurements of a continuous 
decay of the SSS flux until early 2024 April, until only upper limits could be measured.  In other novae, 
 observations of the SSS and its decay indicate
  that the the SSS cooling occurs within about a month 
  \citep[e.g.][]{Nelson2008, Page2022}. Models indicate that the effective temperature should remain constant as long as the burning continues \citep[see, among others][]{Yaron2005} and although the publications of the nova models do not indicate an
   exact duration of the cooling-off period, this is omitted only because it is understood that the SSS is switched off very rapidly, namely in less than a month in most calculations  (Prialnik 2025, private communication). Thus, we probably caught the source in February at the very beginning of cooling-off and the total SSS phase duration is likely to have been close to half a year. This implies a constant bolometric luminosity phase of total duration close to 9 months. According to the models of \citet{2013ApJ...777..136W}, this time is typical of a WD mass of $\sim$1.2 M$_\odot$ (see Fig. 13 of these authors).  The grid of models of these authors has steps of 0.1 M$_\odot$.

$\bullet$ There is additional evidence that the white dwarf in V1716 Sco is likely to have a mass close to 1.2 M$_{\odot}$: our fit the continuum with the atmosphere model yields an effective temperature of 750,000-780,000 K, which is the T$_{\rm eff}$ of a 1.1 M$_\odot$ WD in the models of \citet{2013ApJ...777..136W}, and of a 1.2 M$_\odot$ WD in the models of \citet{Starrfield2012}, again with steps of 0.1 M$_\odot$ in the models' grids.
 This is consistent with other estimates of massive WDs in X-ray luminous novae.

$\bullet$  High resolution grating spectra reveal unprecedented features and a very varied physical scenario in different novae. Because the superimposed
 emission features are not resolved in the broad-band X-ray spectra, they may appear
 to ``raise'' the continuum, causing the derived effective temperature in such CCD-type spectra to be higher. For example, fits to {\sl Swift} XRT spectra yield an effective temperature more than 10\% higher. The {\sl NICER} spectra reveal a little more structure, showing that the atmosphere model is not sufficient for a good fit, but cannot resolve any emission lines in the soft range below 0.7 keV, so that the nature of the additional component remains undetermined. The Spectral paper contains the complete spectroscopic analysis based on our {\sl Chandra} high resolution spectra, yielding information about outflow and the atomic transitions occurring in the shocked plasma of the nova ejecta and circumstellar material.
  
$\bullet$ Clumps in the ejecta, and interactions of the outflow with already emitted circumstellar material cause varying ionization stages in the surrounding medium; this is most likely causing the irregular variability. We found that there is no correlation between the irregular variability and the amplitude of the short, periodic modulations.

 $\bullet$  We note that all the novae and non-nova SSS in which short-period modulations were measured \citep[see][]{Ness2015,Drake2021, Orio2022} had
 an estimated SSS effective temperature exceeding 500,000 K,
   while no periodic modulations were ever measured in the cooler SSS. 
   
$\bullet$ In the analysis of the periodic variability, we found that only long, continuous exposures allow us to rule out period drifts, which are often a simple artifact of the varying amplitude of the pulsations of the SSS flux. A stable period is more likely to be the rotation period of the WD than the result of stellar pulsations, in which an apparent drift may be an effect of overlapping pulsation modes. In the literature, there have been suggestions that the pulsations may be due to non-radial oscillations of the WD
 \citep[e.g.][]{Drake2003, Leibowitz2006}. However, \citet{Wolf2018} and Rosenberg et al. 2025 (preprint) failed to reproduce the
  short $\simeq$35 s oscillations of novae RS Oph or KT Eri as due to non-radial $\epsilon$-mode oscillations. 
    The modes that are excited in the models correspond to periods that are quite shorter than those observed. We argue against the interpretation of an authentic period drift and instead suggest that the varying detected periods are only artifacts of the varying modulation amplitude, particularly when measured in short exposures using the Lomb-Scargle method. Since the main reason to suggest stellar pulsations
 as the cause of the short SSS modulations was indeed the period drift \citep[e.g.][]{Leibowitz2006}, we suggest that instead, assuming a constant period, the WD rotation is the most likely cause
      of the fluctuations. In several cases quoted above, the WD spin has been in fact identified as the cause of other SSS modulation with longer periods (several minutes), in novae that at quiescence have turned out to be IPs. Whether V1716 Sco is also an IP remains to be determined with future follow-up of the nova in quiescence. We note that we still do not have a clear explanation of why the periodicity would be observed in the SSS phase, since the X-ray luminosity of an accretion flow to the poles would be only a very tiny fraction of the SSS luminosity. This fact seems to indicate a non-homogeneous atmosphere of the WD while it is burning in shell, at least with effective temperature $\geq$500,000 K. The lack of correlation of the amplitude of the SSS pulsation with the total SSS flux seems to indicate two
       different origins for the two types of  variability, possibly with the periodic modulation decreasing in amplitude when the
       polar caps are not entirely visible.

\section{Acknowledgments}
This work uses data obtained from the {\sl Chandra} X-ray Observatory, supported by an award to PI Orio, contained in the {\sl Chandra} Data Collection \href{https://cda.cfa.harvard.edu/chaser/searchOcat.action}{DOI: 10.25574/cdc.413}. The {\sl NICER} mission is supported by NASA. We used data software provided by the High Energy Astrophysics Science Archive Research Center (HEASARC), which is a service of the Astrophysics Science Division at NASA/GSFC.
\facilities{Chandra, NICER, Swift, ADS, AAVSO, HEASARC}

\appendix

\section{X-ray Data Tables}
\setcounter{table}{0}
\renewcommand{\thetable}{A\arabic{table}}
\startlongtable
\begin{deluxetable}{lccc}
\tablecaption{Swift Detections and average count rates. PO = Post Outburst \label{tab:Swiftdata}}
\tablewidth{0pt}
\tablehead{
\colhead{\textbf{Obs. ID}} & \colhead{\textbf{Obs. Date}} & 
\colhead{\textbf{Day PO}} & \colhead{\textbf{Net Count Rate (ct \(\mathrm{s}^{-1}\))}}}
\startdata
00015990001 & 2023 Apr 21 & 1.24 & $<$ 0.004 \\
00015990002 & 2023 Apr 25 & 5.62 & $<$ 0.018 \\
00016007002 & 2023 Apr 29 & 9.34 & $<$ 0.023 \\
00016007004 & 2023 May 1 & 11.21   & $<$ 0.084  \\
00016007005 & 2023 May 1 & 11.75  &  $0.016\substack{+0.004 \\ -0.003}$ \\
00016007006 & 2023 May 3 & 12.94       &$0.038\substack{+0.006 \\ -0.005}$     \\
00016007008 & 2023 May 5 & 14.66      & $0.033\substack{+0.005 \\ -0.004}$      \\
00016007009 & 2023 May 7 & 16.70     & $0.093\substack{+0.023 \\ -0.019}$      \\
00016007011 & 2023 May 10 & 20.34      & $0.038\substack{+0.010 \\ -0.008}$       \\
00016007014 & 2023 May 11 & 21.94      &$0.061\substack{+0.007 \\ -0.006}$     \\
00016023002 & 2023 May 19 & 29.14       &$0.061\substack{+0.006 \\ -0.006}$     \\
00016023004 & 2023 May 22 & 32.19      &$0.041\substack{+0.008 \\ -0.008}$   \\
00016023006 & 2023 May 25 & 35.43      &$0.052\substack{+0.007 \\ -0.007}$   \\
00016023008 & 2023 May 28 & 37.61      &$0.044\substack{+0.007 \\ -0.007}$ \\
00016023010 & 2023 May 31 &  41.39      &$0.078\substack{+0.008 \\ -0.008}$     \\
00016066002 & 2023 June 8 & 49.31      &$0.029\substack{+0.005 \\ -0.005}$      \\
00016066006 & 2023 June 10 & 51.30     &$0.046\substack{+0.006 \\ -0.006}$   \\
00016066010 & 2023 June 12 & 53.34      &$0.036\substack{+0.006 \\ -0.006}$     \\
00016066012 & 2023 June 13 & 54.13      &$0.044\substack{+0.006 \\ -0.006}$   \\
00016066014 & 2023 June 14 & 55.46       &$0.052\substack{+0.007 \\ -0.007}$  \\
00015990006 & 2023 June 21 & 62.20        &$0.036\substack{+0.005 \\ -0.005}$   \\
00015990007 & 2023 June 24 & 65.22      &$0.025\substack{+0.005 \\ -0.005}$  \\
00015990008 & 2023 June 27 & 68.45      &$0.034\substack{+0.005 \\ -0.005}$  \\
00015990009 & 2023 June 30 & 70.99       &$0.024\substack{+0.004 \\ -0.004}$  \\
00015990010 & 2023 July 9 & 80.53      &$0.043\substack{+0.019 \\ -0.014}$    \\
00015990013 & 2023 July 18 & 88.84      &$0.065\substack{+0.008 \\ -0.008}$    \\
00015990014 & 2023 July 21 & 92.88       &$0.463\substack{+0.018 \\ -0.018}$    \\
00015990015 & 2023 July 24 & 95.19     &$1.589\substack{+0.061 \\ -0.061}$   \\
00015990016 & 2023 July 27 & 98.03     &$2.049\substack{+0.063 \\ -0.063}$    \\
00015990017 & 2023 July 28 & 98.89      &$0.307\substack{+0.054 \\ -0.054}$    \\
00015990018 & 2023 July 31 & 102.06    &$0.021\substack{+0.004 \\ -0.004}$    \\
00015990019 & 2023 Aug 2 & 103.79    &$0.443\substack{+0.030 \\ -0.030}$   \\
00015990020 & 2023 Aug 3 & 105.24    &$0.374\substack{+0.025 \\ -0.025}$   \\
00015990021 & 2023 Aug 4 & 106.03     &$0.992\substack{+0.035 \\ -0.035}$  \\
00015990022 & 2023 Aug 5 & 107.29    &$0.478\substack{+0.046 \\ -0.046}$  \\
00015990023 & 2023 Aug 6 & 108.07     &$0.143\substack{+0.013 \\ -0.013}$   \\
00015990024 & 2023 Aug 7 & 109.13    &$0.079\substack{+0.010 \\ -0.010}$  \\
00015990025 & 2023 Aug 8 & 109.66     &$0.234\substack{+0.013 \\ -0.013}$    \\
00015990026 & 2023 Aug 9 & 111.11    &$0.931\substack{+0.032 \\ -0.032}$    \\
00015990027 & 2023 Aug 10 & 112.44     &$2.779\substack{+0.055 \\ -0.055}$   \\
00015990028 & 2023 Aug 11 & 112.63    &$2.436\substack{+0.046 \\ -0.046}$   \\
00015990029 & 2023 Aug 12 & 114.15    &$1.418\substack{+0.037 \\ -0.037}$   \\
00015990031 & 2023 Aug 13 & 115.22   &$1.870\substack{+0.045 \\ -0.045}$     \\
00015990030 & 2023 Aug 14 & 115.67    &$1.727\substack{+0.035 \\ -0.035}$    \\
00015990032 & 2023 Aug 15 & 117.46    &$0.033\substack{+0.006 \\ -0.006}$    \\
00015990034 & 2023 Aug 17 & 119.24    &$0.846\substack{+0.039 \\ -0.039}$     \\
00015990036 & 2023 Aug 19 & 121.42    &$0.731\substack{+0.030 \\ -0.030}$     \\
00015990038 & 2023 Aug 21 & 123.20     &$2.058\substack{+0.099 \\ -0.099}$    \\
00015990039 & 2023 Aug 22 & 124.59    &$1.245\substack{+0.060 \\ -0.060}$  \\
00015990040 & 2023 Aug 23 & 124.74   &$1.614\substack{+0.069 \\ -0.069}$  \\
00015990041 & 2023 Aug 24 & 126.19    &$1.324\substack{+0.034 \\ -0.034}$  \\
00015990042 & 2023 Aug 27 & 129.43   &$1.471\substack{+0.060 \\ -0.060}$ \\
00015990043 & 2023 Aug 28 & 130.29  &$1.084\substack{+0.036 \\ -0.036}$  \\
00015990044 & 2023 Aug 29 & 130.74    &$1.301\substack{+0.064 \\ -0.064}$ \\
00015990045 & 2023 Aug 30 & 132.13   &$2.439\substack{+0.064 \\ -0.064}$ \\
00015990046 & 2023 Aug 31 & 132.92  &$1.208\substack{+0.034 \\ -0.034}$  \\
00015990047 & 2023 Sep 1 & 133.79   &$1.533\substack{+0.044 \\ -0.044}$     \\
00015990048 & 2023 Sep 2 & 134.38    &$2.339\substack{+0.060 \\ -0.060}$   \\
00015990049 & 2023 Sep 3 & 136.02    &$0.904\substack{+0.033 \\ -0.033}$    \\
00015990050 & 2023 Sep 4 & 136.88   &$1.533\substack{+0.053 \\ -0.053}$      \\
00015990051 & 2023 Sep 5 & 138.07    &$1.591\substack{+0.043 \\ -0.043}$    \\
00015990052 & 2023 Sep 6 & 139.26    &$2.191\substack{+0.044 \\ -0.044}$   \\
00015990053 & 2023 Sep 7 & 140.25    &$2.439\substack{+0.058 \\ -0.058}$   \\
00015990054 & 2023 Sep 8 & 141.57 &$2.261\substack{+0.056 \\ -0.056}$   \\
00015990055 & 2023 Sep 9 & 141.64    &$2.384\substack{+0.056 \\-0.056}$   \\
00015990056 & 2023 Sep 11 & 144.23   &$2.273\substack{+0.049 \\ -0.049}$    \\
00015990057 & 2023 Sep 14 & 146.83     &$2.794\substack{+0.055 \\ -0.055}$  \\
00015990058 & 2023 Sep 17 & 149.76    &$2.726\substack{+0.062 \\ -0.062}$    \\
00015990059 & 2023 Sep 20 & 153.01     &$3.569\substack{+0.061 \\ -0.061}$    \\
00015990060 & 2023 Sep 23 & 156.38    &$2.850\substack{+0.065\\ -0.065}$    \\
00015990061 & 2023 Sep 26 & 159.08   &$3.711\substack{+0.063 \\ -0.063}$    \\
00015990062 & 2023 Sep 29 &161.86     &$3.006\substack{+0.057 \\ -0.057}$  \\
00015990063 & 2023 Oct 2 & 165.42     &$3.127\substack{+0.055 \\ -0.055}$  \\
00015990064 & 2023 Oct 9 & 172.35    &$3.384\substack{+0.067 \\ -0.067}$   \\
00015990065 & 2023 Oct 16 & 178.63    &$3.786\substack{+0.065 \\ -0.065}$     \\
00015990067 & 2023 Oct 26 & 188.80     &$3.563\substack{+0.062 \\ -0.062}$    \\
00015990068 & 2023 Oct 30 & 192.76   &$2.633\substack{+0.135 \\ -0.135}$     \\
00015990069 & 2024 Feb 1 & 287.16   &$0.320\substack{+0.008 \\ -0.008}$   \\
00015990070 & 2024 Feb 7 & 293.42    &$0.170\substack{+0.011 \\ -0.011}$    \\
00015990071 & 2024 Feb 11 & 296.82    &$0.124\substack{+0.010 \\ -0.010}$    \\
00015990072 & 2024 Feb 15 & 301.10  &$0.068\substack{+0.006 \\ -0.006}$     \\
00015990074 & 2024 Feb 23 & 309.11   &$0.017\substack{+0.003 \\ -0.003}$   \\
00015990075 & 2024 Feb 27 & 312.86    &$0.016\substack{+0.004 \\ -0.004}$   \\
00015990076 & 2024 Mar 2 & 317.17  &$0.012\substack{+0.003 \\ -0.003}$  \\
00015990078 & 2024 Mar 10 & 324.96     &$0.005\substack{+0.002 \\ -0.002}$ \\
0015990078 & 2024 Apr 13 & 358.21 & $<$ 0.010 \\
0015990082 & 2024 Apr 17 & 362.78 & $<$ 0.004 \\
\enddata
\end{deluxetable}
%
\begin{table}
\caption{NICER observations with 
net exposure duration (fitted continuous exposure/net exposure), average count rates, and for continuous exposures longer than 500 s, fitted periods $p_S$ for a sinusoidal (Equation \ref{equation_fit_singlecomp}) and $p_M$ for a multi-component sine function (Equation~\ref{equation_fit_multicomp}) with parameter values of the fit $\phi_1$, $\phi_2$, and $f$, as in the text. Observations with listed exposure times exceeding 500 s but no periodic data are comprised of multiple exposures less than 500 s and the listed exposure time is cumulative. We also give the values of the reduced $\chi^2$ value for the ``simple''  ($\chi^2_{\text{red,S}}$) and for the multi-component fit ($\chi^2_{\text{red,M}}$), respectively and the $\Delta AIC$, computed as $\Delta AIC_S -\Delta AIC_M$. Positive values of $\Delta AIC$ favor the multi-component model (values above 10 indicate a strong preference). Multiple rows are listed for some observations where multiple exposures were individually fit. } \label{tab:NICEROBS}
\begin{center}
\rotatebox{90}{
\begin{tabular}{lcccccccccccc}
\hline
\textbf{Obs. ID} & \textbf{Obs. Date} & \textbf{Day PO} & \textbf{Exposure} & \textbf{Net Count Rate} & \textbf{$\boldsymbol{p_S}$} & \textbf{$\boldsymbol{p_M}$} & \textbf{$\boldsymbol{\phi_1}$} & \textbf{$\boldsymbol{\phi_2}$} & \textbf{$\boldsymbol{f}$} & \textbf{$\chi^2_{\text{red,S}}/\chi^2_{\text{red,M}}$} & \textbf{$\boldsymbol{\Delta AIC}$} \\
& &(days) & (s) & (ct~s$^{-1}$) & (s) & (s) & (rad) & (rad) & & ($\chi^2$/d.o.f.) & \\
\hline
6203910101 & 2023 Jul 25 & 96.84 & 489 & 91.92$\pm{0.71}$ &...&...&...&...&...&...&...\\
6203910102 & 2023 Jul 26 & 97.09 & 547 / 601 & 85.98$\pm{0.65}$ & 71.24 & 75.89 & 2.98 & 1.31 & 0.21 & 1.06 / 1.06 & -0.51 \\
6203910103 & 2023 Jul 26 & 97.99 & 739 / 1,063 & 66.37$\pm{0.49}$ & 75.76 & 81.79 & 4.21 & 4.3 & 0.51 & 1.50 / 1.32 & 120.37 \\
6203910104 & 2023 Jul 28 & 99.35 & 1123 / 2,615 & 11.72$\pm{0.11}$ & 77.74 & 77.82 & 3.28 & 6.28 & 0.11 & 1.29 / 1.25 & 47.11 \\
6203910105 & 2023 Jul 29 & 100.25 & 49 & 6.30$\pm{0.38}$&...&...&...&...&...&...&...\\
6203910106 & 2023 Jul 30 & 101.48 & 992 & 2.32$\pm{0.06}$&...&...&...&...&...&...&...\\
6203910107 & 2023 Jul 31 & 102.38 & 978 & 0.69$\pm{0.05}$&...&...&...&...&...&...&...\\
6203910108 & 2023 Aug 1 & 103.61 & 2,298 & 0.98$\pm{0.04}$&...&...&...&...&...&...&...\\
6203910109 & 2023 Aug 2 & 104.38 & 854 / 1,599 & 29.78$\pm{0.24}$ & 77.89 & 78.19 & 3.38 & 3.72 & 0.09 & 1.30 / 1.25 & 35.92 \\
6203910109 & 2023 Aug 2 &104.56 & 745 / 1,599 & 35.81$\pm{0.22}$ & 77.75 & 77.73 & 4.44 & 4.24 & 0.03 & 1.16 / 1.14 & 2.30\\
6203910110 & 2023 Aug 3 & 105.34 & 770 / 1,718 & 12.51$\pm{0.12}$ & 76.70 & 76.53 & 2.59 & 2.36 & 0.12 & 1.46 / 1.46 & -3.13\\
6203910110 & 2023 Aug 3 & 105.79 & 534 / 1,718 & 15.47$\pm{0.17}$ & 76.92 & 76.92 & 4.13 & 0.01 & 0.38 & 1.06 / 1.03 & 4.71\\
6203910111 & 2023 Aug 5 & 107.54 & 496 & 34.77$\pm{0.35}$&...&...&...&...&...&...&...\\
6203910112 & 2023 Aug 11 & 113.23 & 168 & 58.60$\pm{0.69}$&...&...&...&...&...&...&...\\
6203910114 & 2023 Aug 15 & 117.15 & 841 & 2.20$\pm{0.67}$&...&...&...&...&...&...&...\\
6203910115 & 2023 Aug 16 & 118.06 & 2,720 & 1.33$\pm{0.04}$&...&...&...&...&...&...&...\\
6203910116 & 2023 Aug 18 & 120.57 & 1,746 & 9.47$\pm{0.16}$&...&...&...&...&...&...&...\\
6203910117 & 2023 Aug 19 & 121.03 & 1001 / 1,673 & 46.58$\pm{0.34}$ & 78.99 & 78.85 & 4.44 & 2.30 & 0.21 & 1.19 / 1.12 & 57.12\\
6203910118 & 2023 Aug 19 & 121.99 & 490 & 51.45$\pm{0.46}$&...&...&...&...&...&...&...\\
6203910119 & 2023 Aug 21 & 123.09 & 789 & 76.42$\pm{0.57}$&...&...&...&...&...&...&...\\
6203910120 & 2023 Aug 22 & 124.45 & 1,138 & 47.12$\pm{0.36}$&...&...&...&...&...&...&...\\
6203910121 & 2023 Aug 23 & 125.62 & 560 / 1,668 & 63.90$\pm{0.41}$ & 77.82 & 78.17 & 4.01 & 2.34 & 0.25 & 1.12 / 1.11 & -2.92\\
6203910121 & 2023 Aug 23 & 125.87 & 794 / 1,668 & 55.08 $\pm{26}$ & 77.54 & 77.80 & 2.55 & 6.05 & 0.45 & 1.17 / 1.11 & 37.45\\
6203910122 & 2023 Aug 24 & 126.96 & 158 & 43.33$\pm{0.59}$&...&...&...&...&...&...&...\\
6203910123 & 2023 Aug 26 & 128.45 & 430 & 47.19$\pm{0.45}$&...&...&...&...&...&...&...\\
6203910124 & 2023 Aug 27 & 129.09 & 319 & 30.61$\pm{0.37}$&...&...&...&...&...&...&...\\
6203910125 & 2023 Aug 28 & 130.00 & 397 & 54.72$\pm{0.51}$&...&...&...&...&...&...&...\\
6203910126 & 2023 Aug 29 & 131.62 & 531 & 84.39$\pm{0.65}$ & 80.00 & 80.22 & 5.26 & 4.93 & 0.16 & 1.04 / 1.03 & 1.44\\
6203910127 & 2023 Aug 30 & 132.64 & 377 & 65.01$\pm{0.59}$&...&...&...&...&...&...&...\\
6203910128 & 2023 Sep 2 & 135.09 & 296 & 84.55$\pm{0.74}$&...&...&...&...&...&...&...\\
\hline
\end{tabular}}
\end{center}
\end{table}
\begin{deluxetable}{cccccc}
\tablecaption{Summary of {\sl Chandra} X-Ray Observations: date, average
 count rate and flux in the 15-35 \AA\ range. \label{tab:observations}}
\tablewidth{0pt}
\tablehead{
\colhead{\textbf{Obs. ID}}  & \colhead{\textbf{Obs. Date}} & \colhead{\textbf{Day PO}} & \colhead{\textbf{Exposure  (s)}} & \colhead{\textbf{Count Rate (ct \(\mathrm{s}^{-1}\))}} &
\colhead{\textbf{Flux $\times 10^{-11} \, \mathrm{erg} \, \mathrm{cm}^{-2} \, \mathrm{s}^{-1}$}}
}
\startdata
28048 & 2023 Aug 26 & 128 & 13,104 & 0.520$\pm{0.014}$ &  3.513\\
28496 & 2023 Aug 27 & 129 &  14,210 & 0.587$\pm{0.013}$ & 4.004 \\
28049 & 2023 Oct 20 & 183 & 9,244 & 1.373$\pm{0.021}$ & 9.183 \\
28987 & 2023 Oct 21 & 184 & 8,732 & 1.414$\pm{0.023}$ & 9.422 \\
28988 & 2023 Oct 22 & 185 & 10,768 &1.361$\pm{0.019}$ & 9.114 \\
\enddata
\tablecomments{Exposure length is the  "total on time" minus "dead time". The net (background corrected) count rate is derived from averaging the $\pm{1}$ order event files in the 0.2-2 keV range. The absorbed (measured) flux is obtained by integrating the photon flux over each spectral bin in  the LETG spectrum. The photon flux is obtained in XSPEC by dividing by the effective area (``setplot area'' command).}
\end{deluxetable}
\clearpage
\bibliographystyle{aasjournal}
\bibliography{references}



\end{document}